\documentclass[12pt,preprint]{aastex}

\usepackage{emulateapj5}

\def\kms{km~s$^{-1}$}

\def\gtsim{
\mathrel{\raise.3ex\hbox{$>$}\mkern-14mu\lower0.6ex\hbox{$\sim$}}
}
\def\ltsim{
\mathrel{\raise.3ex\hbox{$<$}\mkern-14mu\lower0.6ex\hbox{$\sim$}}
}
\def\farcs{\hbox{$.\!\!^{\prime\prime}$}}
\def\deg{\hbox{$^\circ$}}

\def\farcs{\hbox{$.\!\!^{\prime\prime}$}}
\def\deg{\hbox{$^\circ$}}

\def\Mch{{\rm M}_{\rm Ch}}
\def\Ca{{\rm Ca}}

\def\Co{{\rm Co}}
\def\DDT{{\rm DDT}}
\def\Fe{{\rm Fe}}

\def\kms{{\rm km} \, {\rm s}^{-1}}
\def\Msun{{\rm M}_{\odot}}
\def\Ni{{\rm Ni}}
\def\Sc{{\rm Sc}}
\def\Ti{{\rm Ti}}

\def\yr{{\rm yr}}

\def\dim#1{{\, \rm #1}}

\slugcomment{Submitted to the Astrophysical Journal }

\shorttitle{HST Images of the SN~1885 Remnant}
\shortauthors{Fesen et al. }
\begin{document}

\title{The Chemical Distribution in a Subluminous Type Ia Supernova: \\ 
       {\sl HST\/} Images of the SN~1885 Remnant\altaffilmark{1} }

\author{Robert A.\ Fesen\altaffilmark{2}, 
        Peter A. H\"oflich\altaffilmark{3},
        Andrew J. S. Hamilton\altaffilmark{4},
        Molly C. Hammell\altaffilmark{2}, 
        Christopher L. Gerardy\altaffilmark{5}, 
        Alexei M. Khokhlov\altaffilmark{6},
        J. Craig Wheeler\altaffilmark{3}
                                            } 
\altaffiltext{1}{Based on observations with the NASA/ESA Hubble Space Telescope,
obtained at the Space Telescope Science Institute,
which is operated by the Association of Universities for Research in
Astronomy, Inc.\  under NASA contract No.\ NAS5-26555.}
\altaffiltext{2}{6127 Wilder Lab, Department of Physics \& Astronomy, 
                 Dartmouth College, Hanover, NH 03755} 
\altaffiltext{3}{Department of Astronomy, University of Texas, Austin, TX 78712; and 
                 Department of Physics, Florida State University, Tallahassee, FL 32306}
\altaffiltext{4}{JILA and the Department of Astrophysical and Planetary Sciences, 
                 University of Colorado, Boulder, CO 80309}
\altaffiltext{5}{Astrophysics Group, Imperial College London, Blackett Laboratory, 
                 Prince Consort Road, London SW7 2BZ}
\altaffiltext{6}{Department of Astronomy and Astrophysics, University of Chicago, 
                 Chicago, IL 60637 }
\begin{abstract}

SN~1885 was a probable subluminous Type Ia supernova which occurred in the
bulge of the Andromeda galaxy, M31, at a projected location 16$''$ from the
nucleus. Here we present and analyze {\sl Hubble Space Telescope\/} images of
the SN~1885 remnant seen in absorption against the M31 bulge via the resonance
lines of \ion{Ca}{1}, \ion{Ca}{2}, \ion{Fe}{1}, and \ion{Fe}{2}. Viewed in
\ion{Ca}{2} H \& K line absorption, the remnant appears as a nearly black
circular spot with an outermost angular radius of $0\farcs40 \pm 0\farcs025$,
implying a maximum linear radius of $1.52 \pm 0.15$ pc at M31's estimated
distance of $785 \pm 30$~kpc and hence a 120~yr average expansion velocity of
$12{,}400 \pm 1400 \, \kms$.  The strongest \ion{Ca}{2} absorption is organized
in a broken ring structure with a radius of $0\farcs2$ (= $6000$ km s$^{-1}$)
with several apparent absorption `clumps' of an angular size around that of the
image pixel scale of $0\farcs05$ (= 1500 km s$^{-1}$).  \ion{Ca}{1} and
\ion{Fe}{1} absorption structures appear similar except for a small \ion{Fe}{1}
absorption peak displaced $0\farcs1$ off-center of the \ion{Ca}{2} structure by
a projected velocity of about $3000$ km s$^{-1}$.

Analyses of these images using off-center, delayed-detonation models suggest a
low $^{56}$Ni production similar to the subluminous SN~Ia explosion of
SN~1986G.  The strongly lopsided images of of \ion{Ca}{1} and \ion{Fe}{1} can
be understood as resulting from an aspherical chemical distribution, with the
best agreement found using an off-center model viewed from an inclination of
$\sim$ 60$^{\circ}$.  The detection of small scale Ca~II clumps is the first
direct evidence for some instabilities and the existence of a deflagration
phase in SNe~Ia or, alternatively, mixing induced by radioactive decay of
$^{56}$Ni over time scales of seconds or days.  However, the degree of mixing
allowed by the observed images is much smaller than current 3D calculations for
Rayleigh-Taylor dominated deflagration fronts.  Moreover, the images require a
central region of no or little Ca but iron group elements indicative for
burning under sufficiently high densities for electron capture taking place,
i.e., burning prior to a significant pre-expansion of the WD.  Using
time-dependent ionization calculations, we show that the presence today of
neutral ions in this 120 yr old remnant can be understood as ejecta
self-shielding from the UV radiation in the M31 bulge.  

\end{abstract}


\section{Introduction}

The `bright nova' S~Andromedae (S~And) was discovered in the central bulge of
M31 (NGC~224) during late August of 1885.  It is believed to have been a Type Ia
supernova (SN~Ia) on the basis of an apparent absence of hydrogen Balmer lines
\citep{CP36,Minkowski39} and a close match of nearly all reported spectral
features to those of SNe~Ia \citep{deVC85}. However relative to normal SN~Ia
events, SN~1885 was subluminous with an unusually rapid initial decline and a
somewhat redder $B - V$ color \citep{deVC85,Patchett85,Graham88,CP88,vdb94}.  A
recent analysis of SN~1885 observations by \citet{vdb02} suggests a decline
rate of $dM_B(15) = 2.2$ mag and a maximum brightness of $M_V = -18.74$ and
$M_B = -17.42$, assuming a M31 distance modulus of ($m - M$)$_{0}$ = 24.4
corresponding to a distance of 760 kpc.

The remnant of SN~1885 remained undetected for over a century despite numerous
searches in the optical and radio
\citep{Baade46,Kraus65,deBruyn73,SB73,DD84,MK84,deVC85,Bou87,Crane92,Eck02}.
The remnant was finally detected in 1989, not through its emission, but as a
small dark spot of absorption in the M31 bulge on an image taken using a
near-UV filter (3900 $\pm$ 100~\AA) sensitive to \ion{Ca}{2} H \& K resonance
line absorption \citep{Fesen89}.  Although \citet{SD01} report a possible weak
radio detection at 8.4~GHz, SN~1885 continues to be essentially unobservable
via emission, with recent deep {\sl Chandra} X-ray imaging of the M31 bulge
showing no associated X-ray source at its position \citep{Kaaret02}.

Following its ground-based detection, optical and UV imaging and spectra of the
SN~1885 remnant (hereafter referred to as ``SNR 1885'') with instruments
aboard the {\sl Hubble Space Telescope} ({\sl HST\/}) revealed a $\approx$
$0\farcs7$ diameter absorption patch produced principally by Ca expanding at
velocities up to 13{,}000~km~s$^{-1}$ \citep{Fesen99}.  The observed low flux
at the \ion{Ca}{2} H \& K blended line center indicated a foreground starlight
fraction of around 0.20, placing SNR~1885 some 65~pc to the near side of the
M31 bulge midpoint, comparable to its projected 55~pc (= $15\farcs6$) distance
from the M31 nucleus.  

Other absorption lines detected in SNR~1885 included \ion{Ca}{1} $\lambda$4227,
\ion{Fe}{1} $\lambda$3441, and \ion{Fe}{1} $\lambda$3720.  A weak imaging
detection of the remnant with the {\sl HST\/} Wide-Field Planetary Camera-2
(WFPC2) in the wavelength range 2200~\AA \ to 3000~\AA, which is rich in strong
\ion{Fe}{2} lines, suggested a depth of absorption consistent with \ion{Fe}{2}
being fully saturated and an iron mass of $0.1 - 1.0$ $\Msun$
\citep{Hamilton00}.  

The fortuitous positioning of a relatively young and nearly freely expanding
SN~Ia remnant silhouetted against M31's bright central bulge offers an unique
opportunity to map the distributions of Ca and Fe rich debris in a SN~Ia.  In
this paper, we present images of SNR~1885 taken with the Advanced Camera
for Surveys (ACS) on {\sl HST} at a significantly higher angular resolution and
signal-to-noise-ratio than previous images.  The ACS images show a sufficiently
well-resolved remnant to distinguish between various proposed SN~Ia explosion
scenarios.  The observations, image results, our modeling results, 
future observations, and conclusions are given in $\S\S2-6$,
respectively.

\section{Observations}

High-resolution optical and ultraviolet images of SNR~1885, detected via
resonance line absorptions of M31 bulge starlight, were obtained between 2004
August and November using the Advanced Camera for Surveys (ACS;
\citealt{Ford98,Pavlovsky04}) aboard {\sl HST}. ACS has two main imaging
configurations and both were used.  The ACS Wide Field Channel (WFC) consists
of two $2048 \times 4096$ CCDs providing a field of view $202'' \times 202''$
with an average pixel size of $0\farcs05$, while the ACS High Resolution
Channel (HRC) consists of a single $1024 \times 1024$ CCD providing a spatial
resolution of $0\farcs028 \times 0\farcs025$ per pixel and a nominal $29''
\times 26''$ field of view.  

The ACS/WFC was used in combination with narrow passband `ramp' filters to
image SNR~1885 in three separate wavelength ranges. These were the resonance
absorption line of Ca~I 4226.73~\AA, the Ca~II H \& K doublet at 3933.66~\AA \
and 3968.47~\AA, and a continuum band free of strong remnant absorption lines
centered at 4600~\AA. Default ACS/WFC and ACS/HRC dither `box' patterns were
used which are 4-point parallelogram patterns with relative pixel coordinates
(0, 0), (5.0, 1.5), (2.5, 4.5), (-2.5, 3.0), thus having both integer and
sub-pixel shifts.

A \ion{Ca}{1} image of SNR~1885 was taken with the 2\% ramp filter FR423N centered at
4227~\AA, resulting in an effective bandpass of 75~\AA \ FWHM and a peak filter
transmission of about 60\%.  Eight 1120~s exposures were taken during four {\sl
HST} orbits on 2004 November 2 in a four-point dither pattern for a
total exposure time of 8960~s.

A \ion{Ca}{2} H \& K absorption image of SNR~1885 was obtained with the ACS/WFC 2\%
ramp filter FR388N centered at a wavelength 3950~\AA, chosen to match the
center of the remnant's broad Ca~II absorption line profile as observed in the
{\sl HST\/} Faint Object Spectrograph (FOS) spectrum \citep{Fesen99}.  At this
wavelength, the filter's bandpass is 100~\AA \ FWHM with a peak transmission of
about 50\%.  Images were taken on 2004 November 8 with a dithering pattern and
exposure times identical to those of the Ca~I images.

A continuum ACS/WFC image was also taken to allow for correcting small-scale
variations in the distribution of background and foreground starlight from the
M31 bulge at the position of SNR~1885.  The continuum image was taken with the
9\% ramp filter FR459M centered at a wavelength 4600~\AA, giving a bandpass of
about 350~\AA\ FWHM and a peak transmission of nearly 80\%.  The spectral
region is adjacent to the \ion{Ca}{1} and \ion{Ca}{2} lines but free of any strong line
absorption.  Due to to the filter's wide bandpass and the increasing luminosity
of M31 bulge stars to longer wavelengths, a relatively short exposure was
adequate and dithered images were taken during just two orbits on 2004 August
25 for a total exposure time of 4666~s.

For obtaining UV \ion{Fe}{1} and \ion{Fe}{2} images, we used the available
broadband filters on the higher resolution ACS/HRC.  For the \ion{Fe}{1}image,
we selected the F330W filter, which peaks at 3300~\AA \ and extends over
2975--3700~\AA.  The filter catches the \ion{Fe}{1} resonances lines at
3021~\AA, 3441~\AA, and 3720~\AA \ \citep{Morton91}, but most of its coverage
is line-free continuum. Consequently, the filter provided only a weak detection
of \ion{Fe}{1}. Indeed, we used the \ion{Fe}{1} image in part as a background
continuum image to help correct the \ion{Ca}{1} and \ion{Ca}{2} images
(\S\S\ref{CaI}, \ref{CaII}).  Dithered F330W images were taken over three
orbits on 2004 October 13 for a total exposure time of 7800~s.

For the \ion{Fe}{2} image, we selected the F250W filter, which peaks in transmission
at 2500~\AA \ but has a broad transmission passband extending over $2250 -
3400$ \AA.  The filter encompasses the \ion{Fe}{2} resonance lines at 2343~\AA,
2373~\AA, 2382~\AA, 2586~\AA \ and 2599~\AA, but it also includes \ion{Fe}{1} lines at
2524~\AA \ and 3021~\AA, the \ion{Mg}{1} line at 2852~\AA, and the \ion{Mg}{2} doublet at
2796~\AA\ and 2803~\AA, all of which are predicted to have significant optical
depths \citep{Fesen99}.  Dithered F250W images were taken over 6 orbits on 2004
November 20 for a total exposure of 15,600~s.

Standard ACS pipeline IRAF/STSDAS\footnote{ IRAF is distributed by the National
Optical Astronomy Observatories, which is operated by the Association of
Universities for Research in Astronomy, Inc.\ (AURA) under cooperative
agreement with the National Science Foundation. The Space Telescope Science
Data Analysis System (STSDAS) is distributed by the Space Telescope Science
Institute.} data reduction was done, including debiasing, flat-fielding,
geometric distortion corrections, photometric calibrations, and cosmic ray and
hot pixel removal.  The STSDAS {\it drizzle\/} task was used to combine
exposures in each filter.

\section{Image Results}
\label{imageresults}

\begin{figure}
\epsscale{0.60}
\plotone{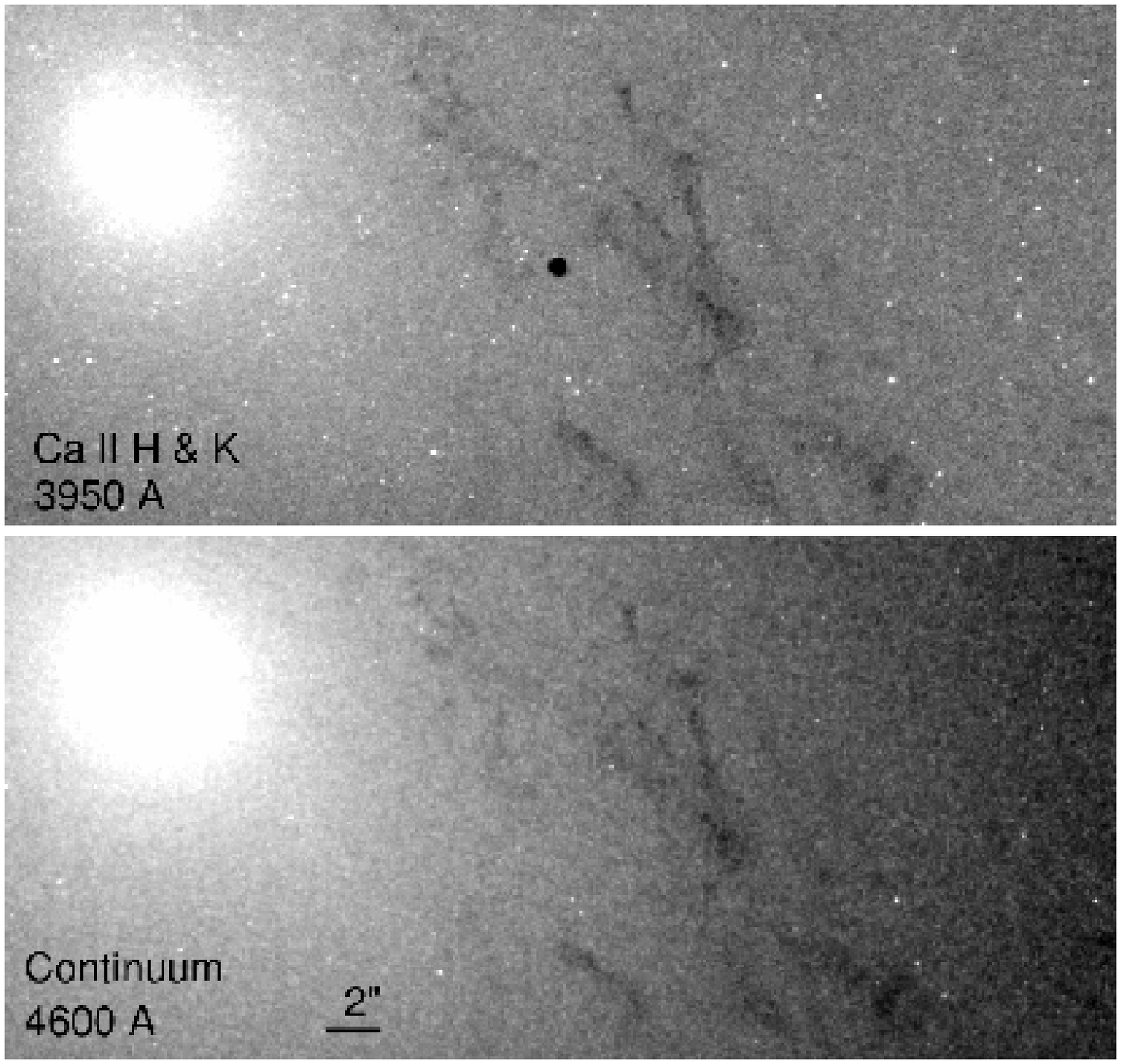}
\caption{{\it Upper panel}: {\sl HST\/} ACS/WFC image of
the M31 bulge centered 16 arcsec southwest of the nucleus.
The ramp filter FR388N centered on 3950~\AA \ (FWHM = 100~\AA) was used to detect the remnant of SN~1885 (S And)
via \ion{Ca}{2} H and K absorption lines (3934, 3988~\AA).
{\it Lower panel}: {\sl HST\/}  ACS/WFC of the same region but taken using the ramp filter FR459M
centered at 4600~\AA \ (FWHM = 350~\AA).
Comparison of these two images shows the presence of the SN~1885 remnant when imaged
in \ion{Ca}{2}. }
\label{fig:Ca2_vs_offband}
\end{figure}

Figure~\ref{fig:Ca2_vs_offband} shows a $40'' \times 20''$ log scaled
intensity view of SNR~1885 and its local environs within the bulge of M31.  The
\ion{Ca}{2} H \& K image ({\it upper panel}) reveals the remnant of SN~1885 as a dark
and circular spot at the historically reported location of SN~1885.  The
coordinates for the center of the remnant are: 
$\alpha(J2000) = 00^{\rm h} 42^{\rm m} 42.96^{\rm s}$, 
$\delta(J2000) = +41{\deg} 16' \, 04\farcs40$.
  
In contrast to the \ion{Ca}{2} image, the offband 4600~\AA \ FR459M continuum
image (Fig.\ 1, {\it lower panel}) shows no appreciable absorption at the
position of SNR~1885. Because SN~1885's projected location is uncontaminated by
the dust lanes that thread the M31 bulge, the lack of appreciable
absorption seen in the 4600~\AA \  image at the position of SNR~1885 suggests
little internal dust formed in this SN~Ia's metal-rich ejecta.

\begin{figure}
\epsscale{0.60}
\plotone{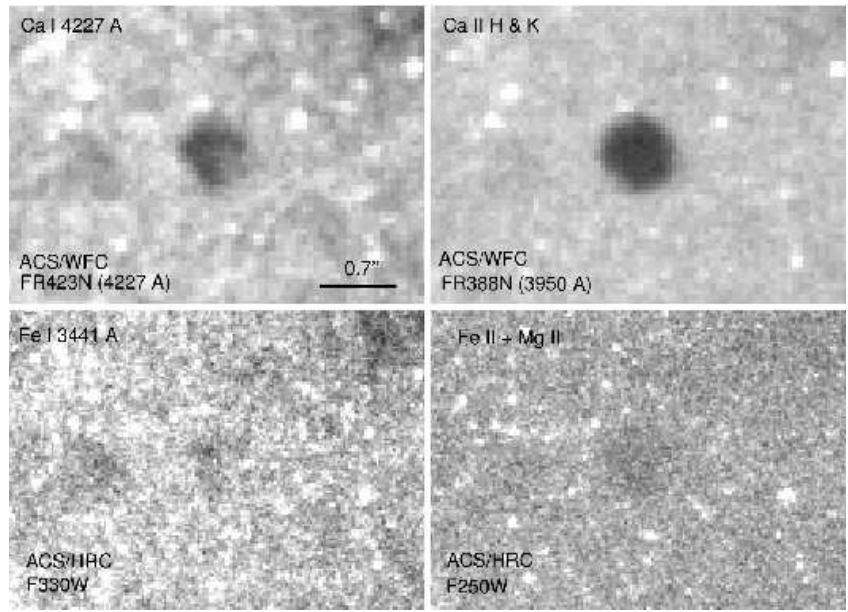}
\caption{{\it Upper panels}: Drizzled ACS/WFC images of the SN~1885 in \ion{Ca}{1} 4227~\AA \
and \ion{Ca}{2} H \& K
at the native scale of $0\farcs05$ per pixel. {\it Lower panels} The remnant's appearance
in \ion{Fe}{1} and \ion{Fe}{2} + \ion{Mg}{2} line absorptions using the ACS/HRC
(image scale: $0\farcs028 \times 0\farcs025$ per pixel) and the
broadband filters F330W and F250W
which are sensitive to \ion{Fe}{1} and \ion{Fe}{2} + Mg II, respectively. }
\label{fig:all_four_species}
\end{figure}

Figure~\ref{fig:all_four_species} shows all four of the individual absorption
images of SNR~1885: \ion{Ca}{1} $\lambda$4227 ({\it upper left}), \ion{Ca}{2} H
\& K $\lambda\lambda$3934,3968 ({\it upper right}), \ion{Fe}{1}
$\lambda\lambda$3441,3720 ({\it lower left}), and \ion{Fe}{2}
$\lambda\lambda$2358,2599 plus \ion{Mg}{2} $\lambda$2800 ({\it lower right}).
The images shown here are in their native spatial pixel scales (ACS/WFC:
$0\farcs05$; ACS/HRC: $0\farcs028 \times 0\farcs025$) and are raw in the sense
that they have not been corrected for any background variations in the surface
brightness of the M31 bulge.

The surface brightness of starlight from the bulge of M31 has both a large
scale gradient and smaller scale fluctuations.  To correct the \ion{Ca}{1} and
\ion{Ca}{2} absorption images of SNR~1885 for these non-uniformities, we used
combinations of the offband 4600~\AA \ continuum image and the `\ion{Fe}{1}'
F330W filter image, in the ratio 4:1 and 3:1, respectively for the \ion{Ca}{1}
and \ion{Ca}{2} images.  As noted above in $\S$2, the F330W filter is broad
and, although it encompasses a number of \ion{Fe}{1} resonance lines, it mostly
covers unabsorbed continuum.  We judged it better to use a background
constructed from continuum both longward (4600~\AA) and shortward (F330W) of
the \ion{Ca}{1} and \ion{Ca}{2} lines, and to accept the mild contamination
from \ion{Fe}{1} absorption in the F330W continuum, than to use the 4600~\AA \
image alone.  Because the ACS/HRC F330W image has a native image scale of
$0\farcs028 \times 0\farcs025$ per pixel, we first resampled the ACS/WFC
\ion{Ca}{1}, \ion{Ca}{2}, and 4600~\AA \ images to $0\farcs025$ per pixel using
the IRAF image interpolation task {\it magnify}, before applying the background
correction.

Figures~\ref{fig:caI_orig_vs_new} and \ref{fig:orig_vs_new} show in close-up
the resampled, background-corrected images of SNR~1885 in \ion{Ca}{1} and \ion{Ca}{2} 
absorption, along with the original, uncorrected images.  We discuss these
images in more detail below in \S\S\ref{CaI} and \ref{CaII}.

The distance to M31 is estimated to be between 760 and 790 kpc based on studies
of M31 Cepheid and RR Lyrae variables, globular clusters, and stars at the tip
of the red giant branch (\citealt{FM90,SG98,Joshi03,Brown04,Rich05,McConn05}
and references therein).  In the discussions below, we have adopted a distance
of $785 \pm 30$ kpc to M31 and the SNR~1885.

\subsection{\ion{Ca}{1}}
\label{CaI}

\begin{figure}
\epsscale{0.90}
\plotone{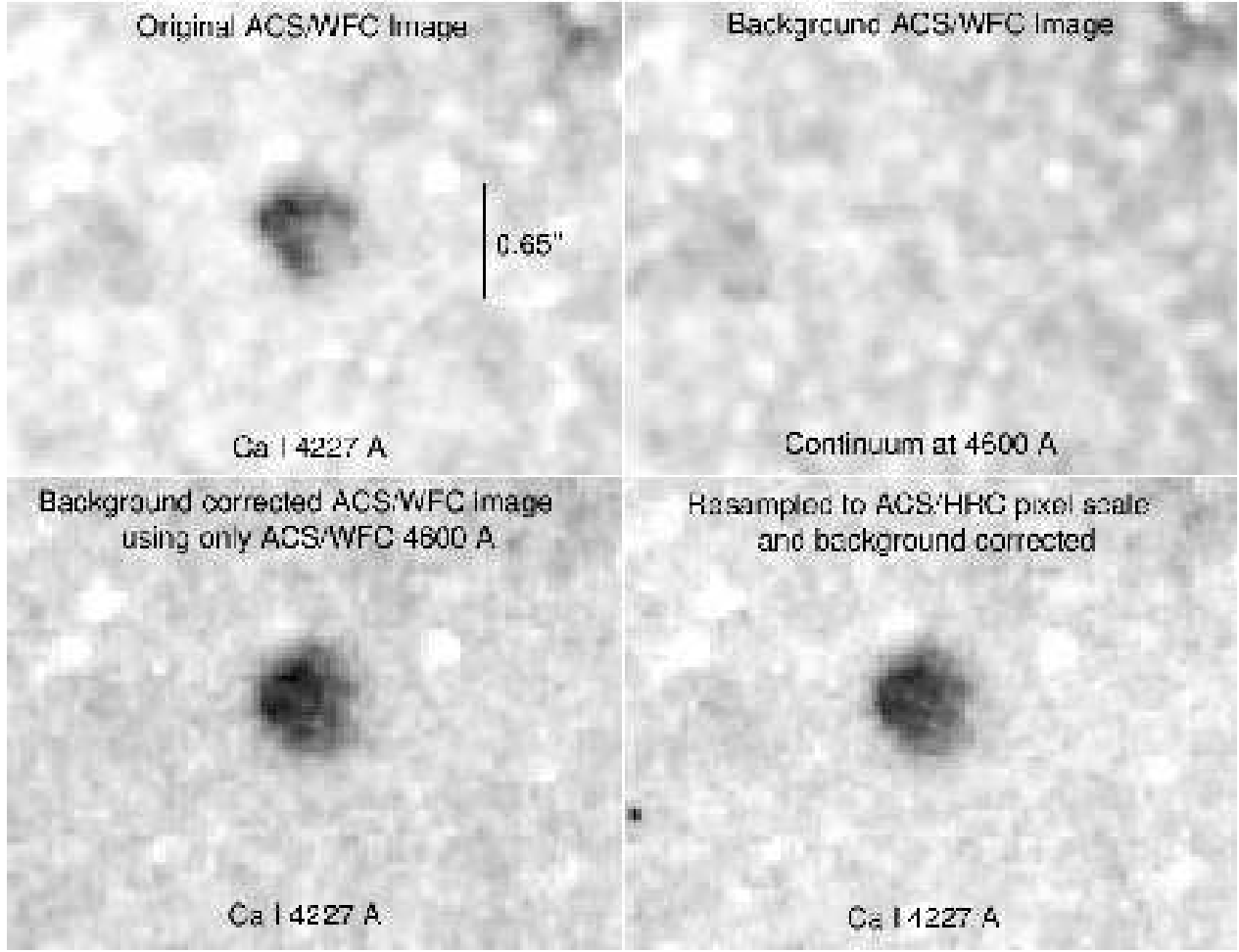}
\caption{Resampled ACS/WFC images of SNR~1885 in \ion{Ca}{1} ({\it upper left}), 4600~\AA \ continuum ({\it upper right}),
         a background corrected \ion{Ca}{1} image resulting from a division of the \ion{Ca}{1} image using the 4600~\AA \ image
         ({\it lower left}), and a \ion{Ca}{1} image background corrected using
         a 4:1 combination of the 4600~\AA \ and the F330W ACS/HRC images ({\it lower right}). See text
         for details. }
\label{fig:caI_orig_vs_new}
\end{figure}

\begin{figure}
\epsscale{0.90}
\plotone{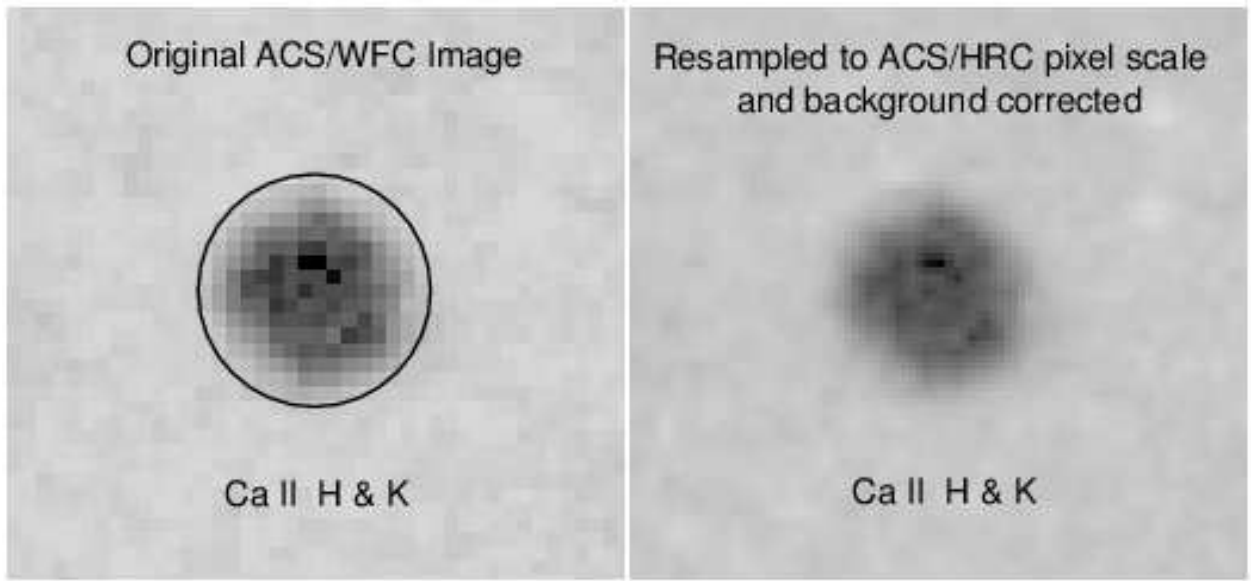}
\caption{Original ACS/WFC \ion{Ca}{2} H \& K image of SNR~1885 ({\it left panel}) and the resampled and
       background corrected \ion{Ca}{2} ({\it right panel}) image
       The black circle shown is $0\farcs80$ in diameter. }
\label{fig:orig_vs_new}
\end{figure}

Figure~\ref{fig:caI_orig_vs_new} shows the \ion{Ca}{1} and 4600~\AA \ continuum images
of SNR~1885 resampled to $0\farcs025$ per pixel.  It also shows two slightly different
background-corrected versions of the \ion{Ca}{1} image.  Like that seen in the raw
\ion{Ca}{1} image presented in Figure 2, the uncorrected but resampled Ca~I absorption image has a
markedly lopsided `C' shaped appearance with the strongest absorption
oriented approximately in the direction of the nucleus of M31 toward the
northeast.  The remnant's overall diameter in this image is roughly
$0\farcs65$ corresponding to a linear diameter of 2.5 pc at a distance of 785
kpc.

The two background-corrected \ion{Ca}{1} images shown in
Figure~\ref{fig:caI_orig_vs_new} are still noticeably lopsided, though much
less so, and both appear more complete in the sense of showing a greater degree
of absorption near the center and along the western limb.  In the first
background-corrected version (Fig.\ 3, {\it lower left}), we used only the
4600~\AA \ continuum image to remove bulge starlight variations, while for the
second version (Fig.\ 3, {\it lower right}) which we regard as being better,
we used a combination of 4600~\AA \ and F330W images for the background in the
ratio 4:1, respectively. One can see that the bulge non-uniformities are
significantly decreased in both corrected images.  Moreover, the good agreement
between the two background-corrected Ca~I images indicates that the mild
contamination from \ion{Fe}{1} absorption in the F330W image does not have a
substantial effect.

While our use of adjacent continuum images should largely remove small scale structures
attributable to variations in the background starlight, we cannot
entirely remove M31 bulge light variations because some 25\% of the bulge starlight
lies foreground to SNR~1885 (see \S\ref{CaII} below). 
Nonetheless, the deepest levels of \ion{Ca}{1} absorptions appear  
along the remnant's eastern and northern limbs.
                                                                                                                                 
\subsection{\ion{Ca}{2}}
\label{CaII}

In the {\sl HST} ACS images, SNR~1885 is most prominently visible through \ion{Ca}{2} 
H \& K line absorption.  Figure~\ref{fig:orig_vs_new} shows the ACS/WFC \ion{Ca}{2} 
image both in its original $0\farcs05$ pixel scale format and then after resampling to
$0\farcs025$ per pixel plus background corrected using the 4600~\AA \ and F330W
images combined in the ratio 3:1. The intensity toward the center of the
uncorrected \ion{Ca}{2} image falls to about 30\% of the background intensity of starlight,
making SNR~1885 the darkest feature in the inner M31 bulge at 3950~\AA.

Both original and background corrected images indicate the likely presence for
a few \ion{Ca}{2} absorption `clumps' having an angular size around that of the
ACS/WFC image resolution ($0\farcs05$), $\sim 0.20$ pc at a distance of 785
kpc. This scale suggests a span in velocity for such clumps of around $1500$ km
s$^{-1}$. The most prominent \ion{Ca}{2} absorption clump is about 25\% darker
than its surroundings and lies some $0\farcs1$ due north of the remnant's
\ion{Ca}{2} absorption center.  While some of the small-scale absorption
structures seen in the \ion{Ca}{2} image are likely due to variations in both
the corrected bulge starlight background and in the starlight lying in front of
the remnant, the strength and non-randomness of the darkest \ion{Ca}{2}
features suggests the existence of a broken ring of \ion{Ca}{2}-rich
clumps at a radius of  $0\farcs2$ ($6000 \, \kms$) located along the inner edge
of a limb-brighten shell.

Aside from these apparent clumps, the degree of \ion{Ca}{2} absorption in the
image appears approximately constant from the center of the remnant out to a
radius of $0\farcs2$, corresponding to a free expansion velocity of $6000 \,
\kms$.  This suggests that the \ion{Ca}{2} absorption is at least partially saturated
in the central region, as previously concluded from the FOS spectrum of
SNR~1885 discussed by \citet{Fesen99}.

The raw and background-corrected \ion{Ca}{2} images of SNR~1885 also indicate a
fairly spherical remnant structure. The
spherical shape and outer extent of the remnant's \ion{Ca}{2} absorption is
illustrated by the  $0\farcs80$ diameter circle shown in Figure 4. This circle
encompasses virtually all of the observed absorption. 

\begin{figure}
\epsscale{0.90}
\plotone{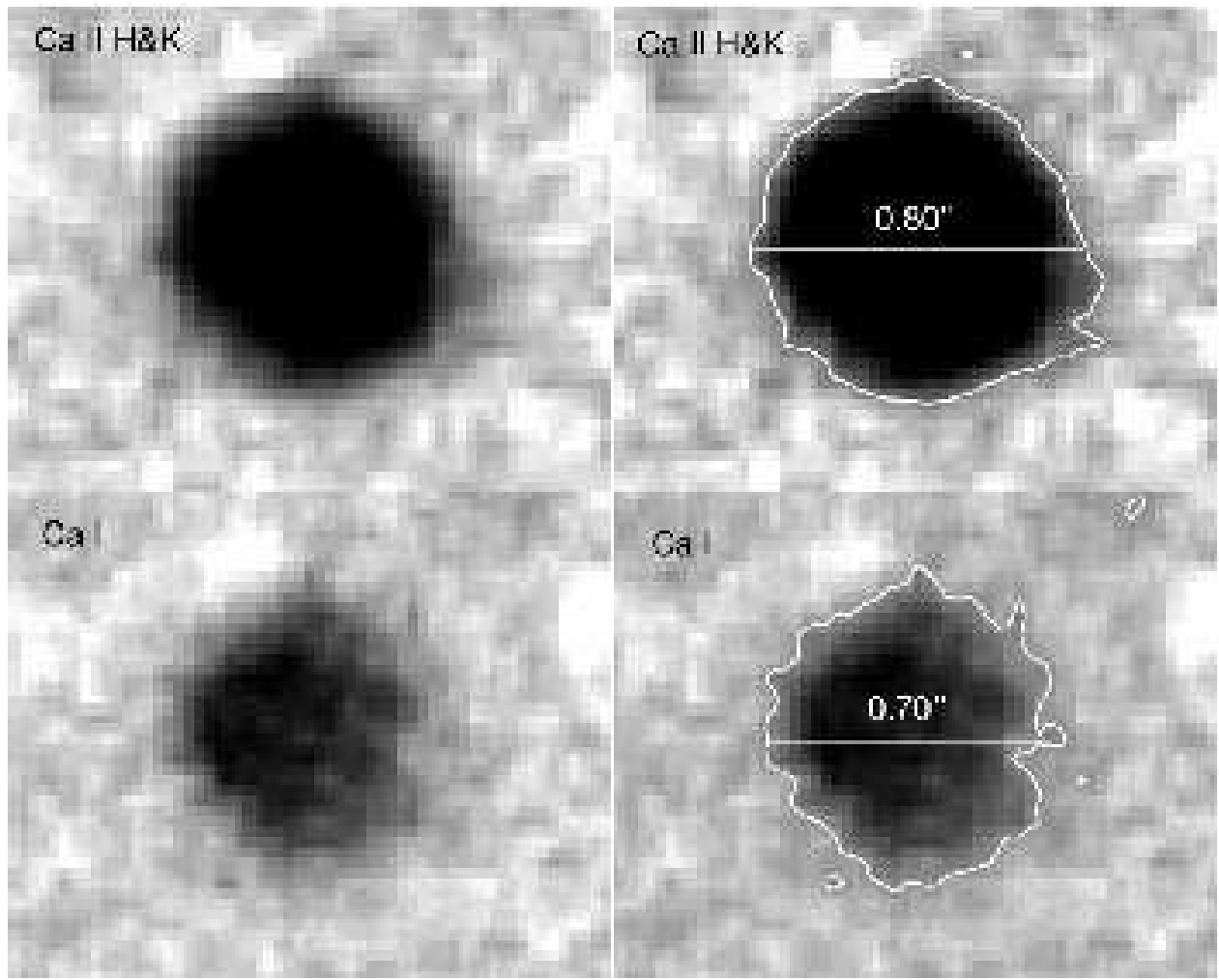}
\caption{Comparison of maximum  \ion{Ca}{2} and \ion{Ca}{1} absorption diameters.
Images are shown scaled as intensity
squared with the contours of 0.85 and 0.97 for \ion{Ca}{2} and \ion{Ca}{1}, respectively.
The \ion{Ca}{2} and \ion{Ca}{1} ACS/WFC images have been corrected for bulge background variations using the
FR459M 4600~\AA \ (FWHM = 350~\AA) image and then 
resampled to $0\farcs025$ to match the higher pixel resolution of the ACS/HRC images.  }
\label{fig:ca1_vs_ca2}
\end{figure}

The maximum detected dimensions of the SN~1885 remnant as seen in both the
\ion{Ca}{2} and \ion{Ca}{1} images are shown in Figure~\ref{fig:ca1_vs_ca2}.
The \ion{Ca}{2} image shows that \ion{Ca}{2} absorption extends out
to a diameter of $0\farcs80 \pm 0\farcs05$, which is larger than that seen via
\ion{Ca}{1} absorption (diameter = $0 \farcs 70$).  Comparison of these
\ion{Ca}{1} and \ion{Ca}{2} images which are shown here greatly stretched to
bring out absorption in the outermost parts of the remnant, shows that the
\ion{Ca}{1} image as a whole appears offset relative to the center of the
\ion{Ca}{2} image, by approximately $0\farcs05$ in the direction of the nucleus
of M31.  The offset is consistent with the asymmetric appearance of the
\ion{Ca}{1} image, and probably reflects a real intrinsic asymmetry in the
distribution of \ion{Ca}{1} versus \ion{Ca}{2}.

The stretched \ion{Ca}{2} image shown in Figure 5 indicates an outer radius for \ion{Ca}{2} 
absorption of $1.52 \pm 0.15 \dim{pc}$ at M31's estimated distance of $785 \pm
30 \dim{kpc}$.  This is equivalent to an expansion velocity of $12{,}400 \pm
1400 \, \kms$ at the $120 \dim{yr}$ age of the remnant.  This velocity
estimate is in agreement with the maximum \ion{Ca}{2} velocity of $13{,}100 \pm
1500 \, \kms$ directly measured from a FOS spectrum obtained by
\citet{Fesen99}.  This agreement supports the notion that the absorbing
\ion{Ca}{2}-rich ejecta are in free expansion.  Moreover, since the image probes the
size in the transverse direction while the spectrum probes the size in the
line-of-sight direction, the similarity of the two expansion velocities
indicates that the \ion{Ca}{2} ejecta structure is approximately spherical.

Finally, as noted above the average central intensity in the \ion{Ca}{2} image
is about $0.30$ that of the local bulge intensity of starlight. This is
somewhat higher than the average central intensity of $0.25$ seen in the FOS
spectrum of SNR~1885 \citep{Fesen99}.  The slightly brighter average central
intensity measured from the ACS/WFC \ion{Ca}{2} image compared to the FOS
spectrum suggests that the foreground starlight fraction may be about $0.25$,
higher than the $0.21$ fraction found by \citet{Fesen99} from a best-fit model
of the FOS spectrum.  A possible reason for this slight discrepancy in central
intensities is that center of the $0\farcs4$ circular aperture of the FOS just
happened to fall on one of the darker patches of \ion{Ca}{2} absorption (see
Fig.\ 4).

\subsection{\ion{Fe}{1}}
\label{FeI}
                                                                                                                        
The remnant's \ion{Fe}{1} absorption distribution was imaged using the broad, near-UV
ACS/HRC filter F330W.  While this filter's bandpass is sensitive to three \ion{Fe}{1} 
lines (3021~\AA, 3441~\AA, and 3720~\AA), the two strongest lines, at 3021~\AA
\ and 3720~\AA, are detected only partially and at low filter transmission
levels; that is, only the redshifted absorption of the 3021 \AA \ line and just
the blueshifted side of the 3720~\AA \ line. In addition, the F330W filter
covered a considerable continuum region between $3000 - 3700$ \AA \ outside of
the \ion{Fe}{1} lines and, in fact, was used to help compose background bulge
starlight images for the \ion{Ca}{1} and \ion{Ca}{2} ACS/WFC images.
                                                                                                                        
Nonetheless, faint \ion{Fe}{1} absorption was detected on the F330W image.  The
\ion{Fe}{1} image (Fig.\ 2) resembles most the uncorrected \ion{Ca}{1} image,
showing a lopsided and broken absorption patch some $0\farcs6$ in diameter, a
break along the southwest limb, and the strongest absorption concentrated
toward the northeast.  The image also shows a peak of \ion{Fe}{1} absorption
roughly $2-3$ pixels across ($0\farcs05 - 0\farcs7$), i.e., about the same size
as that for the \ion{Ca}{2} clumps, displaced toward the northeast by about
$0\farcs1$ from the center of the \ion{Fe}{1} absorption patch.

\subsection{\ion{Fe}{2}}
\label{FeII}

M31 bulge starlight is relatively faint between $2300 - 3500$ \AA \ and
faintest between 2300 -- 2600 \AA \ where several strong \ion{Fe}{2} resonance lines
are found. This fact, together with the rising continuum flux of bulge
starlight toward the redward portion of the $2250 - 3400$ \AA \ passband of the
F250W filter on ACS/HRS resulted in a low signal-to-noise 3000 \AA \
continuum dominated image. This, in turn, produced a very weak detection of
SNR~1885 presumably due to \ion{Fe}{2} line absorption (Fig.\ 2, {\it lower
right panel}) based upon ionization models which predict that the remnant
should be almost black over the $2300 -2600$ \AA \ range \citep{Fesen99}.

Notwithstanding its comparatively poor detection of SNR~1885, the F250W image
bears a clear resemblance to the \ion{Ca}{2} image of the remnant. This can be seen
in Figure 6 were we show a side-by-side comparison of SNR~1885 in \ion{Ca}{2} and the
F250W images.  It is important to note that the F250W filter covers not only
\ion{Fe}{2} lines but also the Mg~II doublet at $2800$ \AA \ and several \ion{Fe}{1} lines
(see $\S$3.3).  However, the similarity of the F250W image to the Ca~II image
but not to the \ion{Fe}{1} or \ion{Ca}{1} images (see Fig.\ 2) suggests that \ion{Fe}{1} does not
make a major contribution to the F250W image.  While \ion{Mg}{2} $2800$ \AA \ could
also contribute to the detection of SNR~1885 on the F250W image, Mg~II is
expected to lie outside of the Ca-rich ejecta region and should thus lead to a
larger diameter appearing remnant. Since this is not what is seen in the F250W image,
we conclude that the F250W image largely shows the distribution of \ion{Fe}{2}-rich
ejecta in SNR~1885. 

\section{Models of SNR~1885}

In this section,
we compare our ACS imaging data of SNR~1885
to off-center delayed-detonation models of Type Ia SNe (SNe~Ia).

\subsection{Overview of SN~Ia scenarios}
\label{scenarios}

There is general agreement that SNe~Ia result from processes
involving the combustion of a degenerate C-O white dwarf (WD) that somehow
reaches the Chandrasekhar mass limit, $\Mch$.  At present, two mechanisms are considered most
likely to bring a WD to this point.  Currently, the leading idea is that a single WD
accretes matter by Roche-lobe overflow from a main sequence or red giant
companion \citep{Whelan73}.  An alternative possibility is the merger of two low-mass
WDs in a binary \citep{Webbink84,it84,pac85,benz90}.
Either way, the explosion is triggered near the center by compression.

One of the main uncertainties in SN~Ia physics is how the nuclear burning flame
propagates through the white dwarf. Specifically, does it propagate as a weak
supersonic detonation front, or as a subsonic deflagration wave?  Whereas
detonation fronts are fairly well understood, modeling the propagation of a
deflagration wave is difficult because the effective speed of deflagration
burning depends on details of the development of multi-dimensional
instabilities which, in turn, depend sensitively on the initial conditions that
lead to the thermonuclear runaway.

Despite significant recent progress, direct deflagration modeling represents a
major computational challenge. This is because of the multi-dimensional physics
involved, the wide range of scales spanning 5 to 6 orders of magnitude, and the
sensitivity to initial conditions.  As a result, fully self-consistent computer
models are beyond reach at the present time, and additional observational
constraints are needed to make progress on the subject.  For example,
observational knowledge of the distribution of the products of nuclear burning,
as in SNR~1885,
can constrain the conditions under which burning must have taken place.

Pure detonation explosions of $\Mch$-mass models have long been ruled out for
normal or subluminous SNe~Ia. Such models predict that the WD is incinerated
almost entirely to $^{56}\Ni$, contrary to observed SN~Ia spectra which show a
range of intermediate mass elements.  Pure deflagration explosions also appear
to be ruled out for normal SNe~Ia.  Three-dimensional numerical computations
show that deflagrations produce deep Rayleigh-Taylor unstable plumes, which
lead to extensive radial, macroscopic mixing of burning products
\citep{Khokhlov01,Roepke03,Gamezo05,Roepke05}.  As a consequence, pure
deflagration models predict a chemically mixed region of burning products,
surrounded by a thick layer of unburned C-O of several tenths of a solar mass.
By contrast, spectra of normal SNe~Ia reveal a chemically stratified structure.
Moreover, there is no evidence for rising deflagration plumes whose signatures
should be visible in infrared spectra of normal SN~Ia at late times or in
subluminous SNe~Ia such as SN~1999by at early times \citep{h02,hetal04}.

Better agreement with observations is obtained with so-called `delayed-detonation' (DD)
models in which the explosion is assumed to begin as a deflagration which
then (somehow) develops into a detonation.  The mechanism by which a deflagration
might turn into a detonation is not well understood.
Thus simulations typically make the {\it ad hoc} assumption that
at some point a deflagration wave simply turns into a detonation,
and the density at which this transition takes place
is treated as a free parameter.

Increasing the transition density
causes the transition from deflagration to detonation to occur earlier,
increasing the amount of $^{56}\Ni$ synthesized.
Variation in this one parameter can
account for the observed gross variation in the properties of SNe~Ia:
how much $^{56}\Ni$ is produced, the overall
distribution of elements, and the brightness of the SN~Ia event
\citep{h95,hkw95,iwamoto99}.

\subsection{Delayed-Detonation Models}
\label{ddchoice}

Arguments about SNe~Ia in general do not necessarily apply to an individual
object such as SN~1885.  However, delayed-detonation models can reproduce
both early to late time light curves and the spectra of both normal and subluminous SNe~Ia
light spectra in the optical to the IR \citep{hetal03}.

We therefore base our comparison to observations of SNR~1885 on a suite of
spherical delayed-detonation models \citep{h02}.  The models start from the
progenitor, 5p01z22, a C-O white dwarf taken from the core of an evolved
5~M$_{\sun}$ main sequence star with solar metallicity.  Through accretion,
this core approaches the Chandrasekhar limit.  A deflagration begins
spontaneously when the core reaches a central density of $2.0 \times 10^9
\dim{g} \dim{cm}^{-3}$ and the WD mass is close to $1.37 \, \Msun$ \citep{h02}.

As mentioned above concerning DD models, the density at which the deflagration is
assumed to transition into a detonation is the key parameter that produces
variation in SNe~Ia.  For extragalactic SNe~Ia, the transition densitycan be 
adjusted so as to match the observed light curve, in particular
its rate of decline \citep{hetal03}.  The chemical profile of the model then predicts a
spectrum that can be compared to observation.  In the case of SNR~1885, we will
instead choose the transition density so as to reproduce the chemical profiles
observed with ACS.  The maximum brightness and light curve then become a
prediction of the model.

The {\sl HST\/} ACS images of SNR~1885 presented in \S\ref{imageresults} show
\ion{Ca}{2} at velocities up to $12{,}400 \, \kms$.  By comparison, DD
models can produce Ca-rich ejecta with maximum velocities that range from $8000 \, \kms$ for a
very subluminous SN~Ia resembling SN~1991bg, up to $14{,}000 \, \kms$ for normal
SN~Ia.  The specific model that best matches SNR~1885 is 5p01z22.16, in which
the `.16' refers to a DD transition density of $16 \times 10^{6} \dim{g}
\dim{cm}^{-3}$ for the progenitor 5p01z22.  In this model, the deflagrated
region encloses $M_\DDT = 0.3 \, \Msun$, and the explosion produces a net total
of $0.27 \, \Msun$ of $^{56}\Ni$.

Model 5p01z22.16, which also provides a good fit to the subluminous SN~Ia
object SN~1986G \citep{Phillips87,Phillips93}, is subluminous by about $0.5$
magnitudes, predicts a peak brightness of $M_B = -18.72$ and $M_V = -18.74$,
and a 15-day decline rate of $\delta m_B = 1.70$ and $\delta m_V = 1.26$.  This
peak brightness in $V$ agrees well with that observed for SN~1885, but the
decline rate is slower than the  $\delta m_B = 2.2$ reported for SN~1885
\citep{vdb02}.  A possible explanation of the discrepancy in the decline rate
may be related to uncertainties in the time $t_{max}$ of maximum light of
SN~1885. All SNe~Ia show a flat maximum, followed by a phase of increasing rate
of decline \citep{phillips99}. A shift of $-3$ days in $t_{max}$ would bring
the model into closer agreement.

\subsection{The SN 1885 Explosion}

In constructing off-center DD models, we have followed the prescription of
\citet{livne99}.  In this prescription, the initial deflagration phase is
modeled assuming spherical symmetry.  The deflagration begins at the center,
and propagates outward in a subsonic deflagration wave.  The energy deposited
by deflagration causes the entire white dwarf to expand.  When the density at
the leading edge of the deflagration wave has fallen to a certain transition
density, a detonation is ignited by hand at a single point (the north pole) of
the deflagration front.  The simulation is then continued in 2 dimensions
assuming cylindrical symmetry.

We choose to impose spherical symmetry on the initial deflagration phase for
two reasons.  The first reason is simply computational tractability:
3-dimensional simulations are computationally expensive, requiring many $10^5$
CPU hours \citep{Khokhlov01,Roepke03,Gamezo05}, and indeed no fully consistent
DD models have ever been calculated.  The second reason is that
multi-dimensional simulations tend to predict mixed chemical profiles, which
are at odds with observations both of typical SN~Ia (see \S\ref{scenarios}
above) and of SNR~1885, where the observed \ion{Ca}{1} image indicates some
radial stratification.  

No mixing seemed to us to offer a more promising
approach than full mixing.  However, in \S\ref{modelimages} below we will show
that the Ca image of SNR~1885 is more filled-in at its center than predicted by
the model, suggesting that the assumption of spherical symmetry over-corrects
the problem, producing more stratification than needed.  We remedy this {\it a
posteriori\/} in \S\ref{modelimages}, by taking the final, fully-exploded WD,
and partially remixing its central regions.

The 3D calculations of \citet{Gamezo05} show that detonation proceeds mainly
along radial fingers, leaving behind connected regions of unburned matter.  To
allow for this, we assumed that the deflagration burns only about $0.9$ of the
matter, leaving the remaining matter unburned.  The unburned material plays an
important role in that it allows the subsequent detonation to move back into
and through the center of the partially deflagrated core, which offers a better
approximation to the 3D calculations of \citet{Gamezo05} than models in which
there is no unburned material, and the detonation front is forced to circulate
around the core \citet{livne99}.

Deflagration burning causes the entire WD to expand.  Several seconds into the
deflagration, when the density has decreased by about two orders of magnitude,
we initiate the detonation at a point on the boundary of the deflagration wave.
We treat the density of the transition as a free parameter, and we characterize
its location by the mass $M_\DDT$ interior to the detonation point.  At the
same time that we initiate the detonation, we switch from spherical to
cylindrical symmetry, following the explosion on a grid of $256 \times 90$
cells in the radial and angular directions.  The detonation begins at the pole
of the cylindrically symmetric configuration, and propagates away from the pole
in all directions.  The detonation front moves not only outward and sideways
through the unburned WD, but also inward through deflagrated core, consuming
the $0.1$ fraction of unburned material there.

In the model, the detonation front reaches the surface of the WD at the pole
about $0.2$ seconds after the detonation, and it reaches the surface at the
opposite pole about $0.4$ seconds after that, consistent with the times
obtained in the 3D calculations of \citet{Gamezo05} (who did not, however,
continue their calculation to homologous expansion).  Over the next 5 to 10
seconds, the hot WD accelerates into free expansion, and the stored thermal
energy is converted into kinetic energy.  The density structure of the
expanding ejecta freezes out as the sound speed plummets.

About one second after entering the
2D phase, we switch back to 1D hydrodynamics, evolving each of the $90$ angular
segments as if it were spherically symmetric.  At this time, angular gradients
in the Ca-rich layers are less than 2\%.

\subsection{Numerical Methods of the Explosion Models}
\label{numerical}

All numerical calculations were done using
the HYDRA code (see \citealt{hh03}).
HYDRA consists of a number of modules that solve simultaneously
the hydrodynamical equations,
nuclear and atomic networks, and radiation transport equations.

For spherical geometry,
the hydro equations are solved in the Lagrangian frame,
and include a front tracking scheme to resolve shock fronts \citep{HWT98}.
For cylindrical geometry,
the coupled hydro and gravitational (Poisson) equations
are solved in an Eulerian frame
using a code based on PROMETHEUS \citep{FAM91}
without adaptive mesh refinement.
The hydro modules use the explicit Piecewise Parabolic Method (PPM)
of \citet{CW84} to solve the compressible reactive flow equations.

HYDRA does not include a detailed model of the nuclear flame
and of the associated development of instabilities.
Instead, the code parameterizes the deflagration speed semi-analytically,
the parameters being adjusted so as to reproduce deflagration speeds
obtained in 3D calculations.
\citet{dominguez2000}
have shown that, in models with a parameterized deflagration speed,
the abundance of chemical elements synthesized
is relatively insensitive to the deflagration speed,
but rather depends mainly on
the amount of electron capture that takes place during deflagration,
which in turn depends on
the total amount of material burned during the deflagration phase.


For the nuclear equation of state and the reactions, HYDRA uses the
nuclear reaction-network library of
\citet[and references therein]{TNH94}.
We take into account 218 isotopes for all models.

For spherical geometry, we use 912 radial layers, while for cylindrical
geometry we adopt a relatively modest resolution of $256 \times 90$ cells.
Computationally, there is a trade-off between a large nuclear network and high
spatial resolution.  We opt for a large nuclear network (rather than the 4 or 5
isotopes typically considered in 3D calculations) because numerical tests
indicate that the final chemical composition is more sensitive to a reduction
in the nuclear network than to a factor 2 reduction in the spatial resolution
\citep{HWT98}.  In any case, the detonation phase is relatively brief and
occurs on an already expanding WD so the effects of moderate spatial resolution
during detonation are not expected to be dramatic.  Still, there is certainly
room for improvement in this aspect of the modeling.

\subsection{The Chemical Distribution in the SNR~1885 Model}
\label{chemical}

Figure~\ref{chem} shows the late-time chemical distribution in model
5p01z22.16, with transition mass $M_\DDT = 0.3$.  This figure shows the
distribution in velocity of O, Si, Ca, Fe, and $^{56}\Ni$ several minutes after
the explosion, which is well into the phase of free, homologous expansion.  The
detonation axis points upward.  Since the $^{56}\Ni$ in SNR~1885 today has long
since decayed radioactively to $^{56}\Fe$, the figure also shows the
distribution of combined $^{56}\Ni + \Fe$.  The chemical structure seen is
similar for models with transition mass $M_\DDT = 0.3$ to $0.5 \, \Msun$.

During the early stage of the deflagration, densities are high enough,
$\ga 10^9 \dim{g} \dim{cm}^{-3}$, that electron capture can take place.
This produces a central region containing stable isotopes of Fe, Co, and Ni
with a mass fraction for Fe of $\approx 50 \%$.
As the density drops in the expanding WD,
electron capture ceases
and the later stages of deflagration yield about $0.1 \, \Msun$
of radioactive Ni and Ca.

The detonation phase produces off-center
layers of explosively synthesized isotopes, from $^{56}\Ni$ and Ca on the
inside to oxygen on the outside.  The element abundances are characteristic of
those obtained in various stages of burning.
The outermost oxygen is partially burned, its mass fraction having increased
from the 50\% of the original C-O WD to about 70\%.

The distribution of Ca in model 5p01z22.16 is qualitatively similar to that seen in
the HST/ACS images of SNR~1885.  The model shows Ca in a lopsided shell with a mass
fraction around 3\% from about 3500 to $11{,}000 \, \kms$, peaking at a mass
fraction of about 6\% at $8200 \, \kms$.  The velocity of the peak increases
slightly to $8600 \, \kms$ if the transition mass is increased to
$M_\DDT = 0.5 \, \Msun$.

The final Fe distribution has an inner region of about 50\%, produced during
the early deflagration. This region is surrounded by a lopsided shell, produced
by detonation to $^{56}\Ni$, where the mass fraction rises to almost 90\%.  The
mass fraction decreases slowly outward, declining to about 1\% at $10{,}000 \, \kms$.

The {\sl HST\/} imaging observations of SNR~1885 reveal not the density itself,
but rather the column density, i.e., the density projected along the
line-of-sight.  Thus Figure~\ref{rho} shows the column densities of Ca and Fe
for Model 5p01z22.16, with the detonation axis taken at various inclinations to
the line-of-sight.  The projected distributions appear round when seen face-on
($0^\circ$) but they become asymmetric at higher inclinations.  The asymmetry
remains modest even at high inclinations, in part because the intrinsic
asymmetry is not large and in part because projection tends to smear out the
asymmetry.  Comparison between model and observed images of SNR~1885 does not
tightly constrain the inclination.  However, the observed \ion{Ca}{1} image is
clearly asymmetric, and we adopt an inclination angle of $60^\circ$ because
this angle is consistent with the projected asymmetry observed.

\subsection{The Ionization Structure of SNR~1885}
\label{ionization}

\citet{Fesen99} and \citet{Hamilton00} argued that the main source of
ionization in SNR~1885 is photoionizing UV from the bulge of M31, and that the
main source of UV opacity in SN~1885 is from resonance lines of neutral and
singly-ionized species.  Recombination and charge exchange are negligible in
SNR~1885 120 years after the explosion.

\citet{Fesen99} also pointed out that if the ejecta were optically thin, then
the lifetimes of \ion{Ca}{1} and \ion{Fe}{1} exposed to UV light from the bulge
would be quite short; $8^{+10}_{-1} \, \yr$ and $10^{+12}_{-1}\,\yr$,
respectively.  The errors quoted come from uncertainty in the line-of-sight
location of SN~1885 within the bulge, but there could be a further factor of two
uncertainty from photoionization cross-sections and from the extinction
correction to the photoionizing UV flux.

Predicted photoionization lifetimes as short as these pose a problem. Although resonance
lines can block about half the photoionizing UV light at the present
time, that still leaves half the UV to get through and reduce the \ion{Ca}{1} and \ion{Fe}{1}
below the currently observed levels.  \citet{Fesen99} suggested
that \ion{Ca}{1} and \ion{Fe}{1} might be present, in spite of their short optically thin
photoionizing times, if the remnant were optically thick in neutral continua,
notably \ion{Fe}{1}, earlier in its history.  They did not, however, carry out the
detailed model computations necessary to demonstrate this possibility.

In the present paper, we have carried out the needed
time-dependent, photoionization computations, and these
show that the observed level of ionization in SNR~1885 is entirely consistent
with model expectations.  We computed the photoionization of SNR~1885 as follows.
We assumed an incident UV spectrum following the {\sl FOS} data of \citet{Fesen99}
down to 2221\,\AA, from the {\sl IUE\/} data of \citep{Burstein88} down
to 1225\,\AA, and from the {\sl HUT} data of \citep{Ferguson93} down to
912\,\AA,  dereddened using the \citet{Cardelli89} extinction curve with color
excess $E_{B-V} = 0.11$ \citep{Ferguson93} and $R \equiv A_V / E_{B-V} = 3.1$.

We treated the photoionizing UV spectrum incident on SNR~1885 as being
isotropic.  \citet{HF91} had noted that SNR~1885 should be receiving more
radiation on the side facing the nucleus of M31 and suggested that for this
reason SNR~1885 should be more ionized on that side.  However, the ACS images
show precisely the opposite.  Not only do the images show \ion{Ca}{1}
concentrated toward the bulge, but an {\sl FOS} spectrum of the remnant showed
\ion{Ca}{1} redshifted by $1100 \, \kms$, corresponding to the side closer to
the bulge \citep{Fesen99}.  Thus an anisotropic photoionization model cannot
account for the observed asymmetry in the \ion{Ca}{1} distribution.  Rather,
the observed asymmetry must reflect an intrinsic asymmetry in the explosion,
such as that produced by an off-center delayed-detonation.  For
simplicity, therefore, we neglected any anisotropy of the incident
photoionization.

The evolution of the ionization structure of C, O, Mg, Si, S, Ca,
Mn, Cr, Fe, and Ni was computed starting from about 10 years after the explosion.  We took
the initial ionization state to be mainly neutral, with an ionization fraction
of 1\% in the central deflagrated region where there is no radioactive
heating, and 10\% in the detonated region where there is radioactive heating,
dominated initially by $^{56}\Ni$ $\rightarrow$ $^{56}\Co$ $\rightarrow$
$^{56}\Fe$ and later by $^{44}\Ti$ $\rightarrow$ $^{44}\Sc$ $\rightarrow$
$^{44}\Ca$.

Although this initial degree of ionization is plausible, it is not
well-determined {\it a priori\/} by the physics because of complications associated
with non-local energy transport by fast electrons in the presence of magnetic
fields \citep{fransson96,l97,ruiz92}.  Tests indicated that the degree of
ionization at the present time is most sensitive to the initial ionization
fraction in the $^{56}\Ni$ detonation layer.  Reducing the ionization in this
layer to 1\% can delay ionization of \ion{Ca}{1} and \ion{Fe}{1} by about 30~years.  The
initial ionization state adopted here was chosen because it is
reasonable and because it reproduces the observed degree of ionization at the
present time.

We computed the temperature structure by taking into account adiabatic expansion,
radioactive heating, and the UV radiation field.  The electron temperature
affects recombination rates and hence the ionization balance at early times
when the density is high.

Line opacities were computed in the Sobolev approximation for homologously
expanding atmospheres, omitting stimulated emission.  In homologously expanding
ejecta, the radius equals velocity times age, $r = v t$, and the column density
per unit velocity $dN/dv$ of any species is equal to its density $n = dN/dr$
times the age $t$. In the nomenclature of \citet{mihalas82},
the Sobolev optical depth $\tau$ at wavelength $\lambda$
for a bound transition with absorption oscillator strength $f$ is
 \begin{equation} \tau = {\pi e^2 \over m_e c} f \lambda n t
\end{equation} where $e$ and $m_e$ are the electron charge and mass, $c$ the
speed of light, and $t$ is the 120\,yr age of the remnant.

Besides resonance lines, we also took into account excited lines with lower
levels originating from the ground level.  We computed the level populations
in a 2-level approximation in which the lower levels are excited by (optically
thin) IR and optical radiation from the bulge and which then
decay by spontaneous emission.  Altogether, we included 6{,}748 bound-bound
transitions with wavelengths, energy levels, and oscillator strengths taken
from \citet{kurucz95}.

Both to follow photoionization and to take into account shielding by neutral
atoms, we included bound-free opacities from the ground state for all neutral
isotopes of the elements above.  Photoionization cross sections were taken from
\citet{verner96} to which we added some resonances treated in the narrow line
approximation based on data from the opacity project \citep{lg95}. Finally, we
also took into account the Doppler shift of the continuum.

The final result of all these calculations is that shielding of UV radiation by
bound-free transitions of \ion{Mg}{1}, \ion{Ca}{1}, and \ion{Fe}{1} proved to
be the key in delaying photoionization so that \ion{Ca}{1} and \ion{Fe}{1} can
survive in SNR~1885 up to the present time.  All the models showed that Ca
and Fe is mostly neutral at early times, but that \ion{Ca}{1} and \ion{Fe}{1} will
rapidly disappear as soon as photoionizing UV radiation can penetrate the
ejecta.  Without bound-free shielding, \ion{Ca}{1} is depleted to 10\% after
about 50 years, whereas with shielding, photoionizing radiation remains blocked
for almost a century.  Whereas the UV opacity from bound-free transitions
evolves substantially, most of the bound-bound transitions remain optically
thick over the life of SNR~1885 up to the present time, so that the UV
line-blocking remains almost constant.

Figure~\ref{ion} shows the evolution of the \ion{Ca}{1}/Ca fraction.  The
high-velocity, outermost layers having a solar-like abundance of Ca are quickly
photoionized, whereas the inner, Ca-rich layers remain opaque for about a
century.  As previously argued by \citet{Fesen99}, Ca-rich ejecta are currently
undergoing a phase of rapid photoionization to \ion{Ca}{2}.  Indeed, our models
predict that \ion{Ca}{1} absorption will become unobservably small in just
another 30 years or so, when the remnant is roughly 150 years old.

\subsection{2D Model Images of SNR~1885}
\label{modelimages}

Figure~\ref{image} shows model \ion{Ca}{2}, \ion{Ca}{1}, and \ion{Fe}{1} images
that can be compared directly to the ACS images.  These images take into
account not only the chemical and ionization structure of the remnant, as
described above, but also the transmission curves of the {\sl HST\/} filters,
the contribution of different lines to absorption, and the finite angular
resolution of {\sl HST}.  As discussed in \S\ref{chemical}, the detonation axis
in the model is taken to be inclined at $60^\circ$ to the line-of-sight.

The model predicts the \ion{Ca}{2} image to be insensitive to the precise
ionization structure, since Ca is mostly singly-ionized.  The model also
predicts that the Ca image should have a sharp edge, where the mass fraction of
Ca drops rapidly from 3\% to the pre-explosion, solar-like abundance.  The
images shown in Figure~\ref{image} are for model 5p01z22.16, which as described
in \S\ref{ddchoice} was chosen so as to reproduce the maximum velocity of
\ion{Ca}{2} observed in SNR~1885.

The \ion{Ca}{2} model image in Figure~\ref{image}
is as seen through the {\sl HST} ramp-filter FR388N
centered at $3950$ \AA and $100$ \AA wide,
and takes into account the doublet character of the \ion{Ca}{2} H \& K absorption.
The model predicts that the \ion{Ca}{2} optical depth should exceed unity
in a ring from $3100$ to $5500 \, \kms$,
which are close to the values measured for SNR~1885.

The model \ion{Ca}{1} and \ion{Fe}{1} images in Figure~\ref{image}
appear off-center by about $3000$ and $2000 \, \kms$, respectively,
again in good qualitative agreement with the observed images.
The offset is caused primarily by
the asymmetric chemical distribution produced by the assumed off-center
delayed-detonation.  The asymmetry is enhanced by the self-shielding of denser
regions against photoionization.

A defect of the model is that the predicted \ion{Ca}{1} and \ion{Ca}{2} images
are emptier at their centers than observed in SNR~1885.  It is natural to
attribute this defect to our assumption that the early deflagration phase of
the explosion was spherically symmetric, which artificially suppresses radial
mixing by Rayleigh-Taylor instabilities in the deflagration wave.  Indeed, the
observed images show structure down to scales $\sim 1500 \, \kms$ comparable to
the angular resolution of the observations, such as might perhaps be produced
by instabilities.

To remedy this defect of the model,
we tried radially re-mixing the final chemical profile in various ways.
Figure~\ref{mix} shows the effect on the model \ion{Ca}{1} and \ion{Ca}{2} images
of radially remixing of:
(a) layers within the central deflagrated region;
(b) layers within the outer boundary of the Ca layer; or
(c) the entire remnant.
All three cases fill in the central Ca hole in the model to a greater or lesser degree.

The observed Ca images of SNR~1885 are intermediate between cases (a) and (b),
suggesting that there may have been a moderate amount of radial mixing between
the inner deflagrated region and the surrounding detonated layers.  This level
of mixing is less than that predicted by current 3D calculations of
RT-dominated deflagration fronts \citep{Khokhlov01,Roepke05}.  Smaller amounts
of mixing could result if the initial conditions at the onset of runaway
burning produce a less violent deflagration \citep{livne05}, or if the WD
undergoes one or more pulsations with mixing prior to the transition to
detonation \citep{hoeflich04}.  Mixing may also be produced or enhanced by
heating of $^{56}\Ni$ rich regions, well after the thermonuclear explosion, on
time scales of hours to days.  Both mechanism are compatible with fluctuations
on scales observed.

\section{Alternative Scenarios for SNR~1885}
\label{alternative}

Because of the known diversity of SNe~Ia, below we speculate on some alternative
scenarios for modeling the specific case of SNR~1885.

A variation of `classical' DD models
(which assume {\it ad hoc\/} that deflagration converts into a detonation),
are the so-called `confined-detonation' models
recently suggested by \citet{Calder03} and \citet{plewa04}.  Their proposal is
motivated in part by the extreme sensitivity to initial conditions of the early
phase of burning.  In this model, a buoyancy driven individual plume rises from
the central deflagrating region to the surface, where it triggers a He-detonation.
A detonation wave then propagates inward, exploding the WD.  The
confined-detonation model may offer a natural mechanism to burn a WD partly via
deflagration and partly by detonation.

Current models of confined-detonation use only a limited nucleosynthetic network,
so there is more work to be done,
but the results to date suggest that
burning in the central layers occurs at densities well below the value of
$10^{9} \dim{g} \dim{cm}^{-3}$
\citep{plewa04}
needed to produce a central region with electron capture, such as is inferred from SNR~1885.
Moreover, the confined detonation models predict a significant amount of Fe
at high velocities, which is not seen in normal SNe~Ia.

Mergers may contribute to the SNe~Ia population and are obvious candidates for
off-center explosions. For S-Andromeda, merger scenarios will suffer from the 
same problem as confined detonations namely, a low central density with negligable 
electron capture, unless the primary star is close to
$\Mch$ when the detonation is triggered. However, a detonation in a $\Mch $
mass WD would produce a large amount of Ni and consequently an over-luminous
SNe~Ia which is incompatible with observations of the SN~1885 event.

Finally, one might even consider SN~1885 as a possible core-collapse supernova.
The fact that the remnant is freely expanding out to $13{,}000 \, \kms$
indicates that the remnant has so far swept up less than about $1.4 \, \Msun$
of material, implying an ambient circumstellar or interstellar density less
than about $3 \, \dim{cm}^{-3}$.  This relatively low density, the historical
record, and the absence of evidence of recent star formation in the bulge of
M31, all point to a SNIa event, as opposed to a SNIb or SNIc event in which a
massive star undergoes core collapse after having lost its hydrogen envelope in
a wind.  Our ACS observations add further evidence in this direction, the
abundance and distribution of Fe and Ca being more consistent with a
deflagration/detonation event than with He-rich freeze-out.  The near spherical
symmetry of the \ion{Ca}{2} image can also be construed as supporting the SN~Ia
interpretation, since there is growing evidence that the explosion mechanism of
these objects is intrinsically asymmetric with axis ratios of 2:1 or more
(\citealt{wang01,leonard01}).

\section{Future Observations of SNR~1885}
\label{future}

The remnant of SN~1885 offers an unique and remarkable opportunity to
investigate the explosion dynamics and chemistry of a SN~Ia explosion in some detail.
The {\sl HST\/} ACS data on SNR~1885 presented in this paper
provide only a glimpse of the potential to probe SN~Ia physics.
In this light, we thought it useful to consider future observations
that might provide further insights.

Figure~\ref{uvtime}
shows a model UV absorption spectrum at three different times,
respectively 90, 120, and 140 years after the explosion.
The predicted spectrum is rich with strong, broad absorption features from dozens of
resonance lines many of which evolve in strength fairly rapidly.

Of course, the relative faintness of the bulge of M31 in the UV will limit
the observability of the SNR~1885 spectrum.
However, useful observations can be made even with limited signal-to-noise.
For example, as discussed in \S\ref{ionization},
the remnant is undergoing a period of rapid ionization of
\ion{Ca}{1} and \ion{Fe}{1} to the singly-ionized state,
so that UV continuum absorption by neutral Ca and Fe
is currently decreasing rapidly.
This reduction in continuum opacity could be detected by broad-band UV imaging
over a period as short as 5~years.

Future spatially resolved UV spectral imaging,
similar perhaps to that available with
the Space Telescope Imaging Spectrograph ({\sl STIS}),
could provide a significant step forward in our understanding of SN~Ia events.
For example, Figure~\ref{uvtiling}
shows UV spectra taken with slits at various offsets from the remnant's center.
Spectra such as these would provide a goldmine of information about the
3-dimensional structure and composition of the supernova ejecta,
and hence about uncertain aspects of explosion physics in SNe~Ia.

While narrow slits would of course provide the most precise mapping of ejecta,
even broad slit spectra would provide useful information.
Figure~\ref{uvslit}
shows predicted UV spectra taken with various different slit widths
available using the STIS on
{\sl HST}.

\section{Conclusions}
\label{summary}

SN~1885, exploding as it did so close to, and to the near side of, the nuclear bulge
of M31, presents a remarkable opportunity to map the chemical and density
structure of freely-expanding ejecta in a Type~Ia supernova.  In this paper, we
present high angular resolution {\sl Hubble Space Telescope} images of the remnant of
SN~1885 sensitive to \ion{Ca}{1}, \ion{Ca}{2}, \ion{Fe}{1}, and \ion{Fe}{2} resonance line absorption.

The highest quality image is that of \ion{Ca}{2}, which shows a spherical
remnant $0 \farcs 80 \pm 0 \farcs 025$ in diameter, corresponding to a
free-expansion radius of $12{,}400 \pm 1{,}400 \, \kms$.  Although the image is
partially saturated (optical depth $\sim$ a few), it shows indications of a
ring-like interior structure peaking at a radius of about $6000 \, \kms$, with
a few apparent absorption `clumps' with an angular size near $0\farcs05$
($1500 \, \kms$).  The SN~1885 remnant appears somewhat lopsided and smaller
(dia.\ = $0 \farcs 65 \pm 0 \farcs 05$) in \ion{Ca}{1} than \ion{Ca}{2},
indicating that the inner parts of the remnant are more neutral than the outer
parts, as expected if the ejecta are being photoionized by UV light from the
bulge of M31.  The remnant's \ion{Ca}{1} and \ion{Fe}{1} absorption structures appear
similar in size and morphology, except that the \ion{Fe}{1} image
shows a small \ion{Fe}{1}
absorption peak of diameter $0\farcs05$ displaced to the northeast
$0\farcs1$ (projected velocity of $3000 \, \kms$) from the center of the
\ion{Ca}{2} structure.  Finally, a weak detection of SNR~1885 across the
wavelength range of $2300 - 3400$ \AA is probably caused by near saturated
absorption from several resonance \ion{Fe}{2} lines.
The \ion{Fe}{2} image appears similar in size to that seen in \ion{Ca}{2}.

We argue that the off-center and ring-like characteristics of the Ca and
\ion{Fe}{1} images favor off-center delayed-detonation models of the explosion.
We considered alternative scenarios including
pure deflagration models, confined-detonations, and WD mergers,
but find all of these to be less likely.

Using the HYDRA code \citep{hh03}, we have made detailed
computations that follow delayed-detonation models of SN~1885 from
the beginning of nuclear runaway up to the present 120~year age of the remnant.
The models follow in detail the nucleosynthesis in, radiative transfer through,
and ionization structure of, the ejecta, including the effects of late-time
photoionization by UV light from the bulge of M31.

We selected a best-match model so as to reproduce the observed
maximum velocity of Ca.  The best-match model undergoes a transition from
deflagration to detonation at a mass fraction of $M_\DDT = 0.3$. The model is
subluminous by about $0.5$ magnitudes, consistent with indications from the
historical record of SN~1885.  The model synthesizes $0.27 \, \Msun$ of
$^{56}\Ni$, and has a light curve similar to that of the subluminous Type Ia
event, SN~1986G.

We find that our model is able to reproduce many of the general observational
features of SNR~1885.  Ca and Fe in SNR~1885 are mostly singly-ionized at the
present time, so the \ion{Ca}{2} and \ion{Fe}{2} images trace essentially all the Ca and Fe
in the remnant.  By contrast, \ion{Ca}{1} and \ion{Fe}{1} trace only the regions with the
highest density and abundance of Ca and Fe, because only in these regions are
\ion{Ca}{1} and \ion{Fe}{1} sufficiently self-shielding against photoionizing UV from the
bulge of M31 to survive up to the present time.

Our analysis presents the first direct comparison between the properties
of a supernova remnant and multi-dimensional explosion models
to probe the underlying physics of burning fronts.
Within the framework of delayed-detonation (DD) models, the
deflagration-detonation transition occurs at densities of the order of
$10^{7} \dim{gm} \dim{cm}^{-3}$,
consistent with the Zel'dovich mechanism  \citep{khokhlov97,n97}.
Potential candidate
processes for mixing of burned and unburned matter may be instabilities
internal to the deflagration front \citep{khokhlov97,n97}, during a pulsational
phase of the WD \citep{khokhlov03,hkw95}, or externally induces by shear
instabilities present in rapidly, differentially rotating WDs when rising
plumes enter this region of strong differential rotation
\citep{hetal03,uenishi03,yoon04a,yoon04b}.

The apparent presence of small scale \ion{Ca}{2}-rich clumps is the first
direct evidence for some instabilities and thus the existence of a deflagration
phase in SNe~Ia or, alternatively, mixing induced by radioactive decay of
$^{56}$Ni over time scales of seconds or days.  However, the degree of mixing
allowed by the observed images is much smaller than current 3D calculations for
Rayleigh-Taylor dominated deflagration fronts.  This could indicate a serious
problem in the current generation of multi-dimensional models for deflagration
fronts, which may be related to the initial conditions at the time of the
runaway.  In addition, the images require a central region of no or little Ca
but iron group elements indicative for burning under sufficiently high
densities for electron capture taking place, i.e., burning prior to a
significant pre-expansion of the WD. 

Off-center, delayed-detonation models have the virtue of being able to reproduce
light curves and spectra of normal and subluminous SNe~Ia \citep{hetal03}.  It
is remarkable that these same models are able to reproduce the distribution and
morphology of Ca and Fe in the remnant of SN~1885.  However, it should be
emphasized that the models described in this paper are parametrized, and thus
not fully self-consistent. Our efforts should be regarded only as a first step
toward tightening the link between explosion physics and
observations of young supernova remnants.

The details of SNe~Ia remain poorly understood \citep{h05}. How the
deflagration begins and develops is extremely sensitive to initial conditions
\citep{hs02,Livne04,Gamezo05}.  Equally unclear is how the deflagration
subsequently turns into a detonation. While observations of any one object are
unlikely to completely resolve such difficult questions, further observations
of the SN~1885 remnant might yield additional insights and guidance for future
computer simulations of SN~Ia explosions.

\acknowledgments This work was supported by NASA through grant GO-10118 to RAF,
MCH, and PAH from the Space Telescope Science Institute, which is operated by
the Association of Universities for Research in Astronomy.  PAH is also
supported by the NSF grant AST-0507557.  CLG is supported through UK PPARC
grant PPA/G/S/2003/00040 and JCW by NSF AST-0406740.

\clearpage
\newpage

%

\begin{figure}
\epsscale{0.90}
\plotone{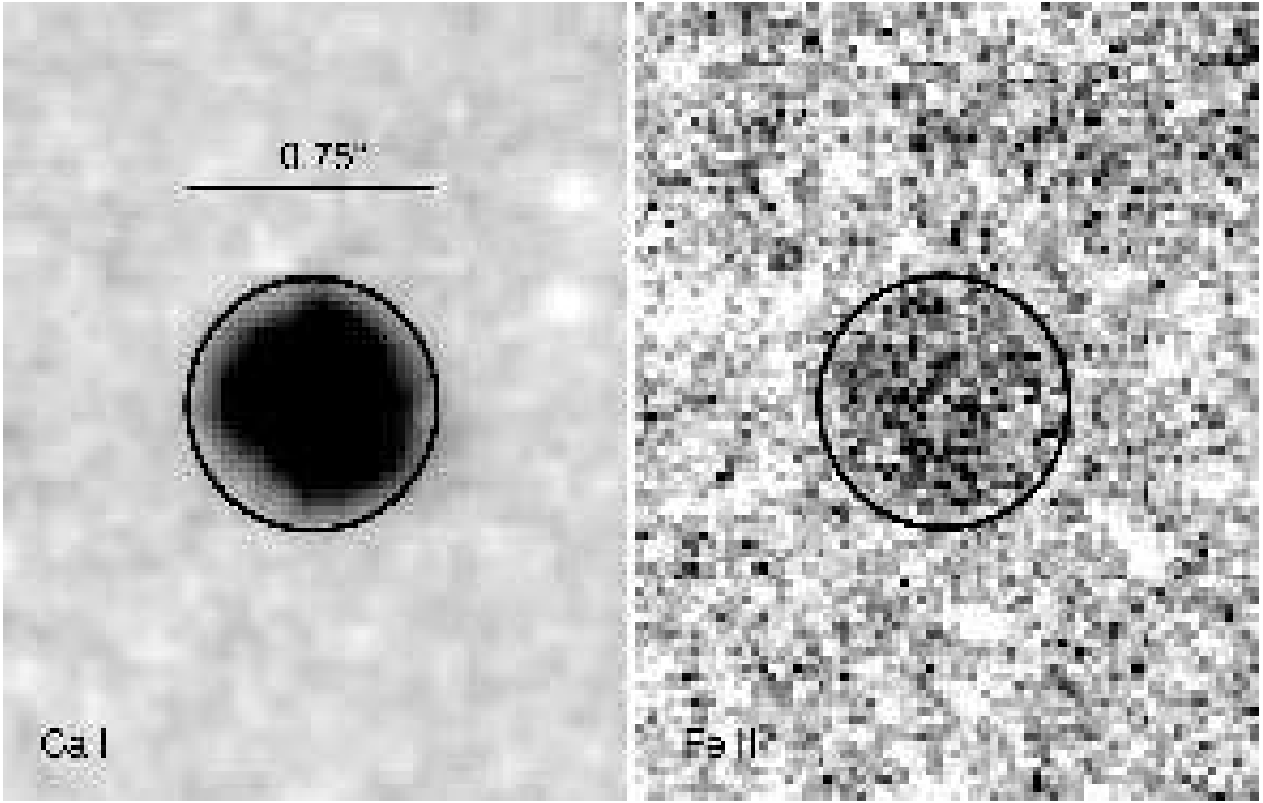}
\caption{Comparison of \ion{Ca}{2} absorption with the \ion{Fe}{2} ACS/HRC image taken with
          the F250W filter sensitive to \ion{Fe}{2} absorption. }
\label{fig:ca2_vs_fe2}
\end{figure}

\begin{figure}
\epsscale{0.90}
\plotone{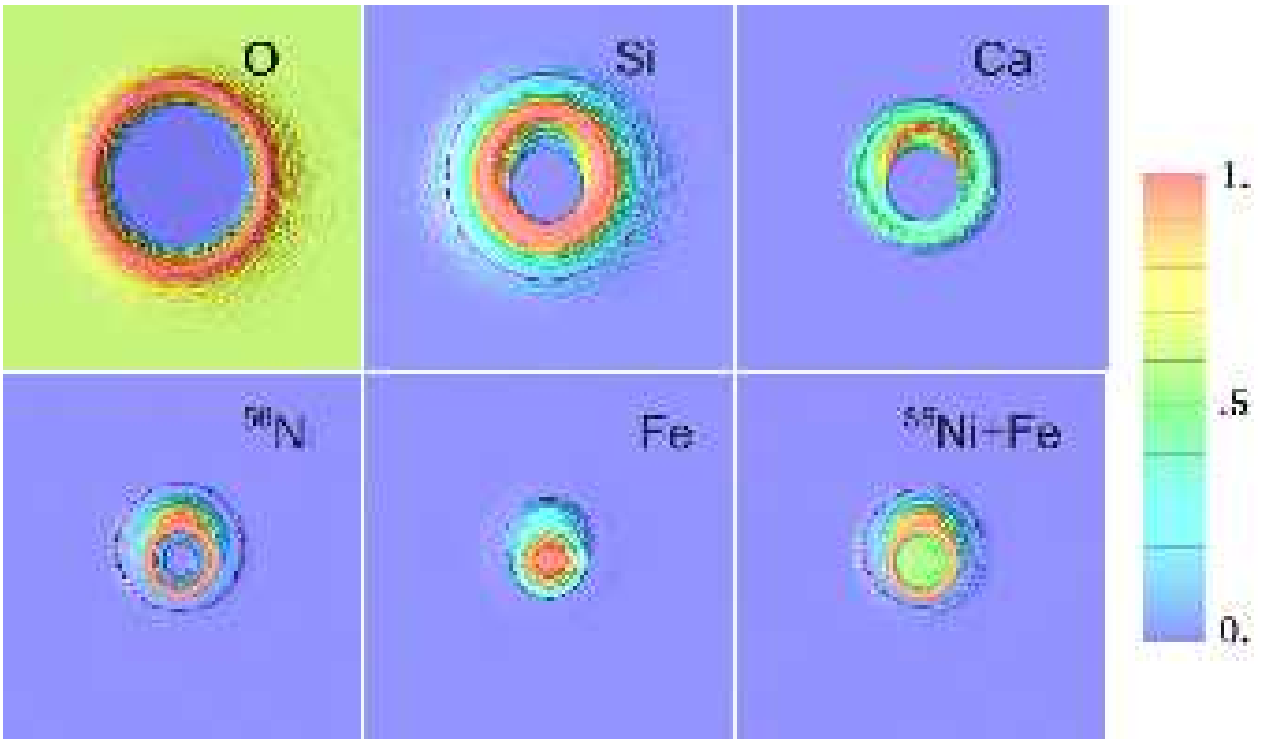}
\vskip -0.0cm
\caption{Chemical distribution of $^{56}Ni$ and stable iron as a function of velocity for the
off-center DDT at $0.3 \, \Msun$. Boxi dimensions correspond to velocities from $-27,500$ to 27,500~km~s$^{-1}$.
The abundances are color-coded from zero (blue) to the maximum (red). The maximum mass fractions
of O, Si, Ca, stable Fe and radioactive $^{56}Ni$ are 0.69, 0.65, 0.061, 0.59 and 0.90, respectively.
In addition, we show the total Fe distribution about 120 years after the explosion.
The central region consists of stable iron group elements
Mn, Fe, Co, and Ni with mass fractions of $\approx 4, 2, 58 $ and $ 36 \%$, respectively
(see H\"oflich  et al. 2002). The color bar is normalized to the maximum
and used in all figures hereafter. }
\label{chem}
\end{figure}

\begin{figure}
\epsscale{0.90}
\plotone{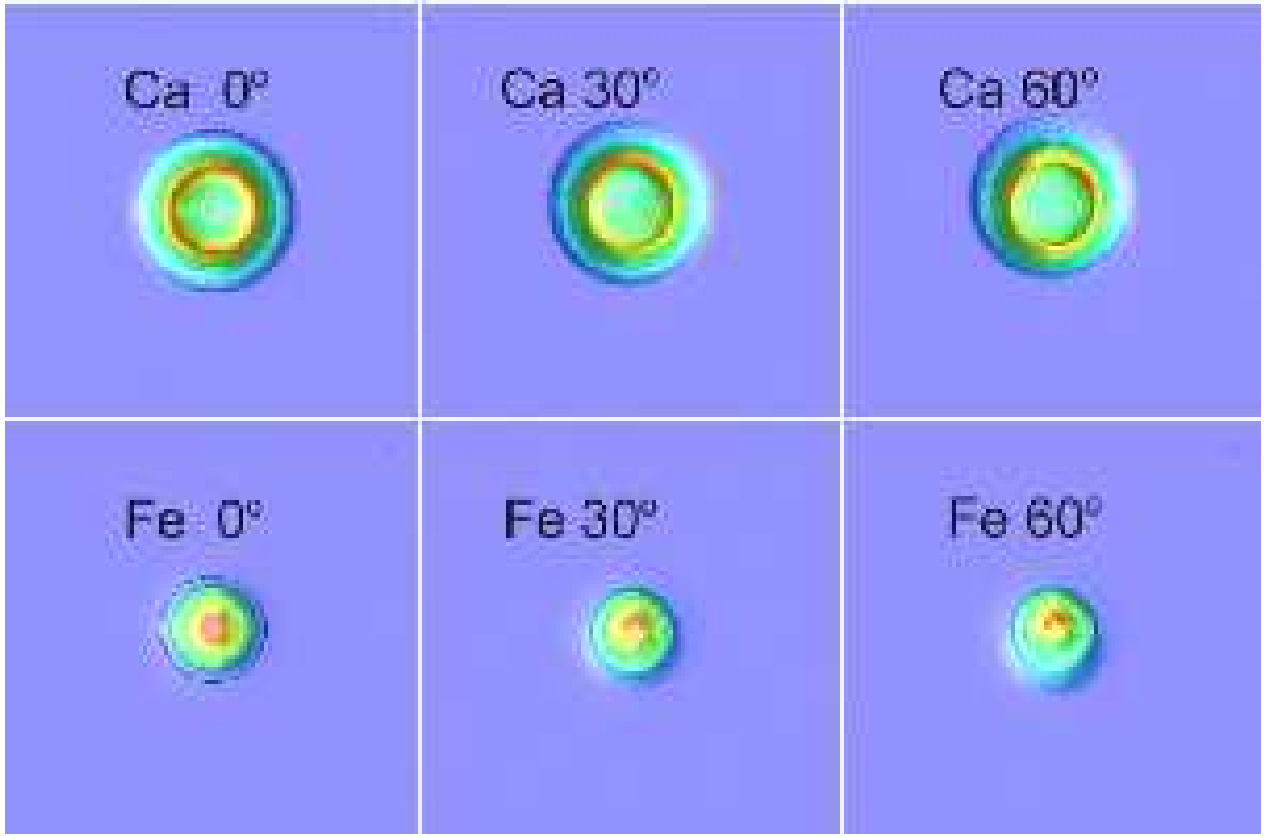}
\vskip -0.0cm
\caption{Column depths for the convolution of the density and chemical profiles for Ca and
Fe as seen from various inclinations for the reference, off-center DD-model.
Overall, the images are quite round.
If viewed pole-on ($90^o$), the chemical structure for Ca is a continuous ring which starts
to break-up at lower latitudes.
}
\label{rho}
\end{figure}

\begin{figure}
\includegraphics[width=4.6cm,angle=270]{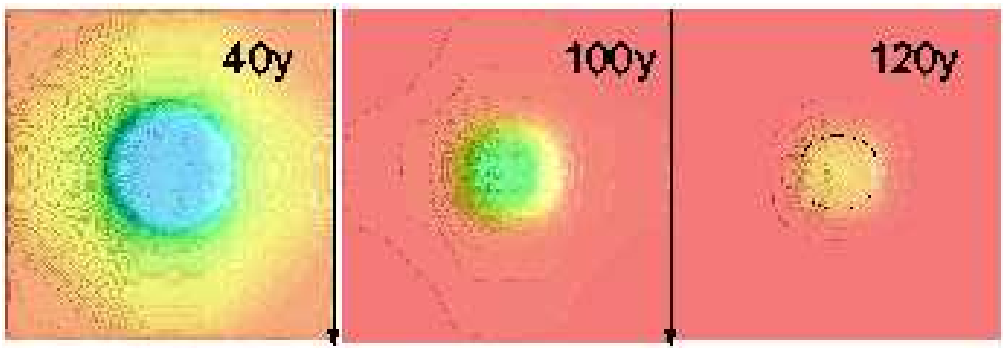}
\vskip -0.0cm
\caption{\ion{Ca}{1}/Ca as a function at year 40, 100 and 120 with a central value of
85\%, 45\%, and 16\%, respectively. The ionization structure of Fe is rather similar
in the models but, due to shielding by \ion{Ca}{1}, larger by $\approx 3 \%$.
Note that there is no \ion{Ca}{1} inside the 0.1 contour at day 120. }
\label{ion}
\end{figure}

\begin{figure}
\epsscale{0.90}
\plotone{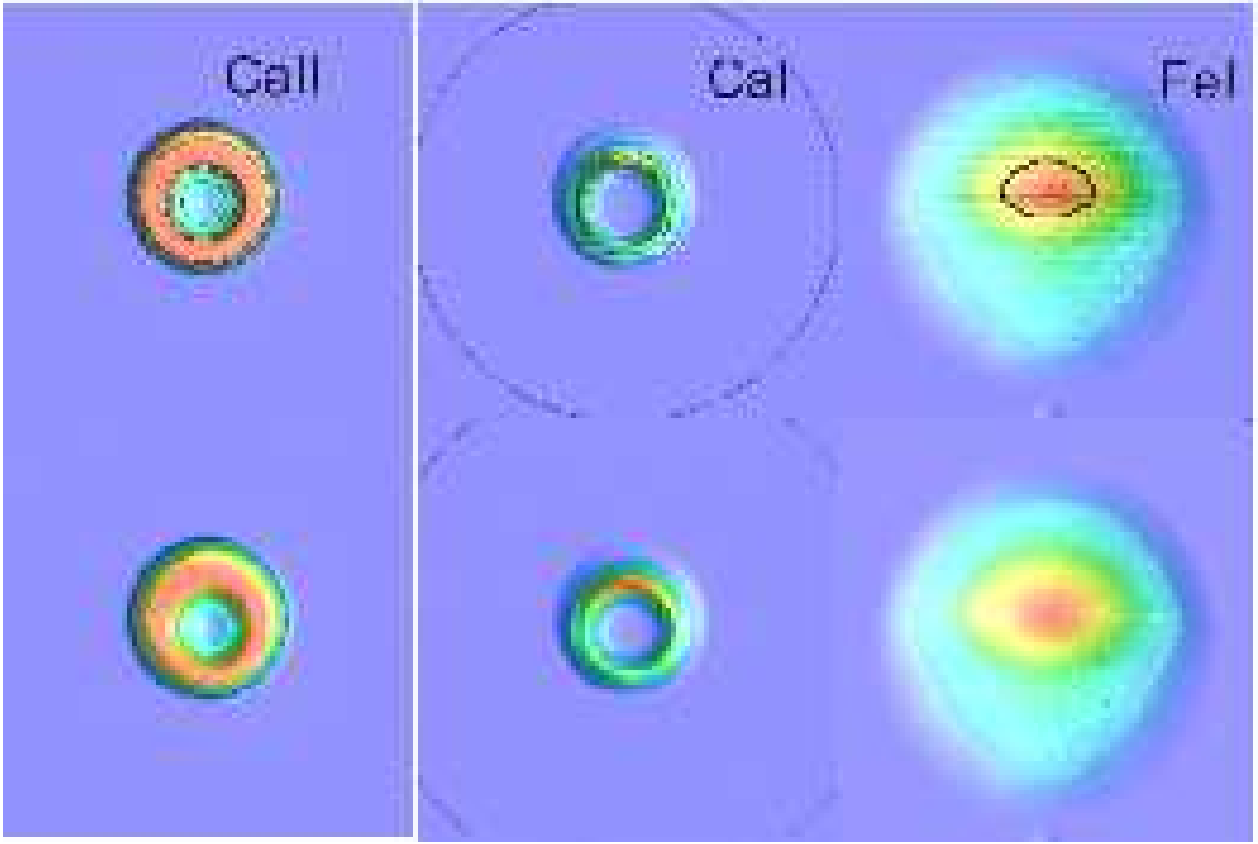}
\vskip -0.0cm
\caption{{\it Upper panels:} Theoretical images \ion{Ca}{2}, \ion{Ca}{1}, and \ion{Fe}{1}
images of the off-center models seen at an inclination of $60^o$.
The contours mark the layers with a darkness of 0, 0.25, 0.50 and 1.00. For \ion{Ca}{1}, \ion{Ca}{2},
and \ion{Fe}{1}, the maximum
darknesses are 0.64, 1.00 and 0.6, respectively. {\it Lower Panels:}
Theoretical images convolved with the HST \ion{Ca}{2}, \ion{Ca}{1}, and \ion{Fe}{1} filter response
functions.  Note that additional line blends tend to smear out the central hole due to the shift in frequency space.
}
\label{image}
\end{figure}

\begin{figure}
\epsscale{0.90}
\plotone{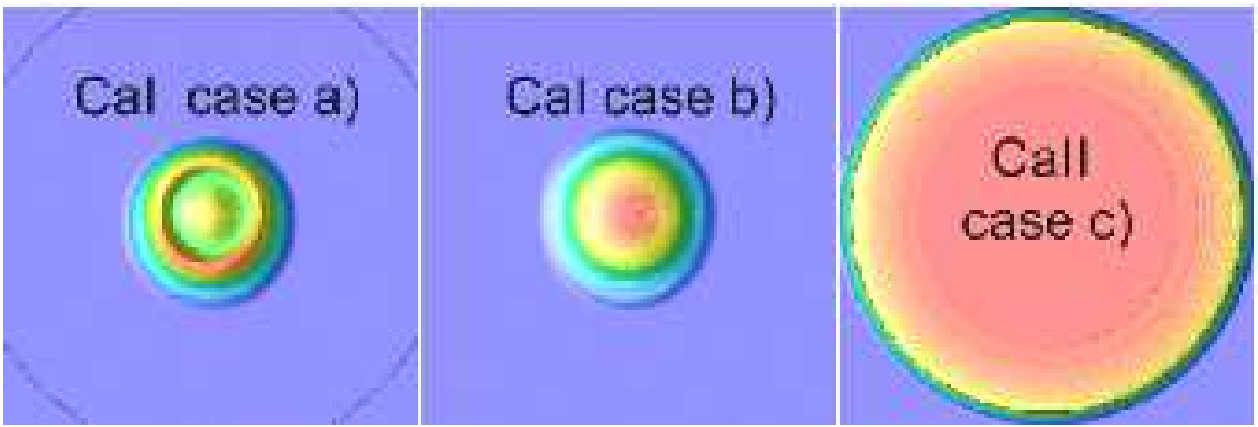}
\vskip -0.0cm
\caption{Influence of mixing on the \ion{Ca}{1} and \ion{Ca}{2} images in case of (a) mixing layers
produced during the deflagration, (b) all layers within the Ca-shell, and (c) all layers
that have undergone burning as expected for a (hypothetical) pure deflagration
which burns the entire WD. For \ion{Ca}{1}, case (b) and (c) are similar because Ca is fully
ionized at the outer layers.
}
\label{mix}
\end{figure}

\begin{figure}
\plotone{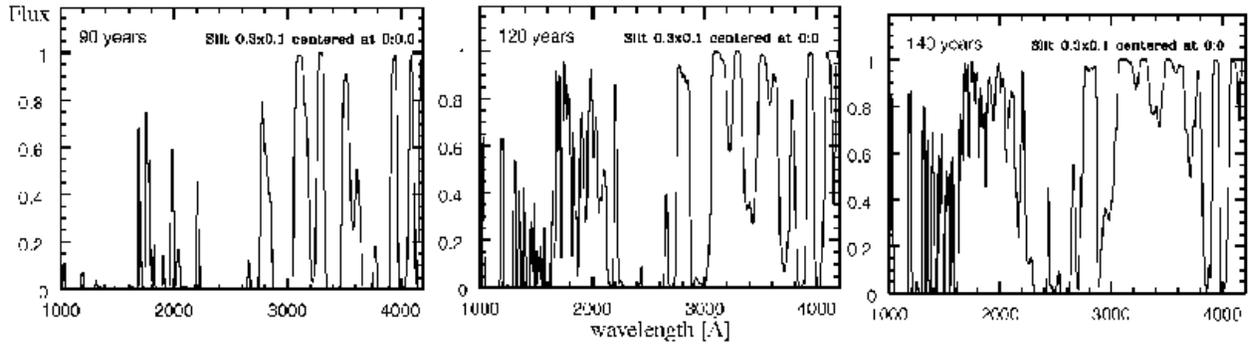}
\vskip -0.0cm
\caption{Predicted UV spectra at 90, 120 and 140 years after the explosion
with a spectrograph of a slit dimension of $0 \farcs 1 \times 0 \farcs3$  similar to that
available on STIS.
The slit has been assumed to be centered and aligned along the axis of symmetry defined
by the off-center DDT and center of the white dwarf.
Note the rapid spectral evolution below 2500~\AA.}
\label{uvtime}
\end{figure}

\begin{figure}
\plotone{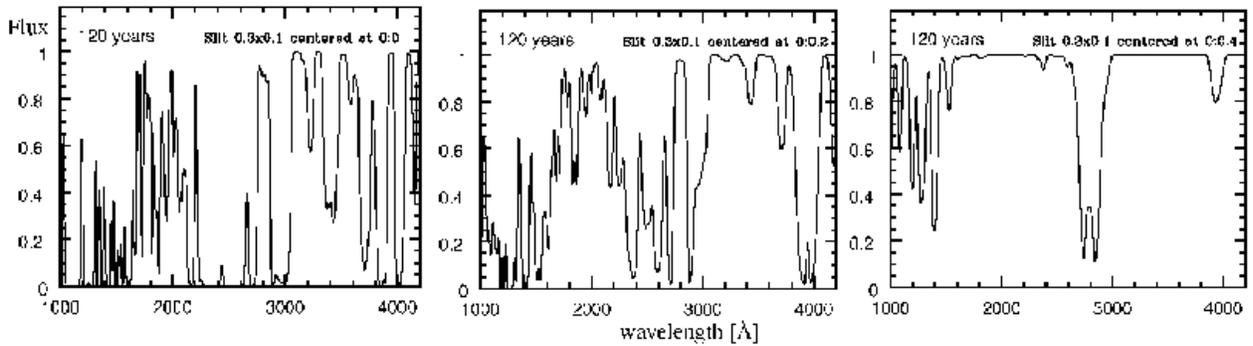}
\vskip -0.0cm
\caption{Same as Figure \ref{uvtime} for 120 yr but now with various slit off-sets relative to
remnant center. }
\label{uvtiling}
\end{figure}

\begin{figure}
\epsscale{0.80}
\plotone{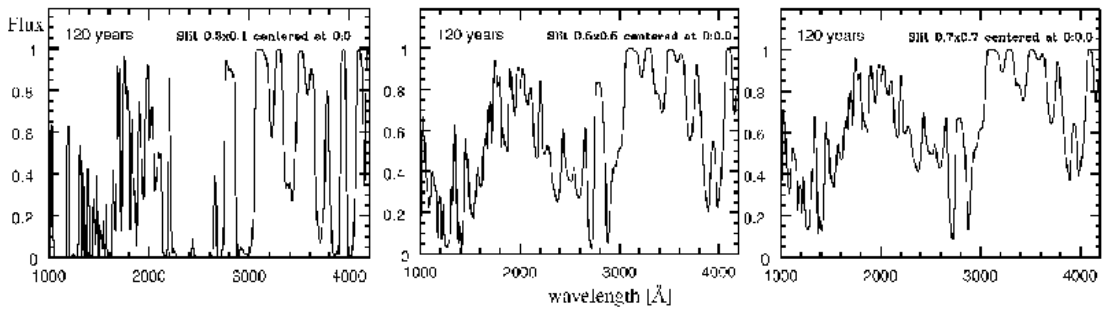}
\vskip -0.0cm
\caption{Predicted UV spectra as in Figure \ref{uvtime} for 120 yr but now as a function of the
slit size. }
\label{uvslit}
\end{figure}

\end{document}